\documentclass[a4paper,12pt]{article}

\usepackage{amsfonts}
\usepackage{amsmath}
\usepackage{amssymb}
\usepackage{latexsym}
\usepackage{color}
\usepackage{colordvi}
\usepackage{epsfig}
\usepackage{array}
\usepackage{pifont}

\newcommand{\beq}{\begin{eqnarray}}
\newcommand{\eeq}{\end{eqnarray}}

\newcommand{\Lams}{\Lambda_{\overline{\rm MS}}}

\newcommand{\be}{\begin{equation}}
\newcommand{\ee}{\end{equation}}
\newcommand{\lwrsim}{\raise0.3ex\hbox{$<$\kern-0.75em\raise-1.1ex\hbox{$\sim$}}}
\newcommand{\lgrsim}{\raise0.3ex\hbox{$>$\kern-0.75em\raise-1.1ex\hbox{$\sim$}}}

\def\C2#1#2{({\cal C}_2)_{#1}^{#2}}

\def\eq#1{eq.~(\ref{#1})}

\def\VEV#1{\langle #1 \rangle}

\def\prd#1#2#3{Phys.\ Rev.\ {\bf D#1} (#2) #3}
\def\npb#1#2#3{Nucl.\ Phys.\ {\bf B#1} (#2) #3}
\def\plb#1#2#3{Phys.\ Lett.\ {\bf B#1} (#2) #3}

\usepackage{geometry}
\usepackage{amsxtra}
\usepackage{axodraw}
\usepackage{cite}

\newcommand{\tr}{\text{Tr}}
\newcommand{\re}{\mathbb{R}\text{e}}

\newcommand{\ghvertex}{\begin{picture}(100,25)(0,-3)
\SetWidth{1.2}
\DashArrowLine(12.5,0)(50,0){5}
\DashArrowLine(50,0)(87.5,0){5}
\Gluon(50,0)(50,25){-4}{3}
\CCirc(50,0){5}{Black}{Yellow}
\Text(12.5,-10)[l]{k}
\Text(87.5,-10)[r]{q}
\Text(60,20)[l]{q-k}
\end{picture}}
\newcommand{\ghost}{\begin{picture}(150,25)(0,0)
\SetWidth{1.2}
\DashArrowLine(12.5,0)(37.5,0){5}
\DashArrowLine(37.5,0)(112.5,0){5}
\DashArrowLine(112.5,0)(137.5,0){5}
\SetWidth{1}
\Vertex(37.5,0){2}
\Vertex(112.5,0){2}
\GlueArc(75,0)(37.5,0,90){-4}{6}
\GlueArc(75,0)(37.5,90,180){-4}{6}
\CCirc(75,37.5){10}{Black}{Blue}
\end{picture}}
\newcommand{\gluonTwoA}{\begin{picture}(150,25)(0,0)
\SetWidth{1.2}
\Gluon(12.5,0)(37.5,0){-4}{2}
\Gluon(37.5,0)(112.5,0){-4}{6}
\Gluon(112.5,0)(137.5,0){-4}{2}
\SetWidth{1}
\Vertex(37.5,0){2}
\Vertex(112.5,0){2}
\GlueArc(75,0)(37.5,0,90){-4}{6}
\GlueArc(75,0)(37.5,90,180){-4}{6}
\CCirc(75,37.5){10}{Black}{Blue}
\end{picture}}
\newcommand{\gluonTwoB}{\begin{picture}(150,25)(0,0)
\SetWidth{1.2}
\Gluon(12.5,0)(75,0){-4}{6}
\Gluon(75,0)(137.5,0){-4}{6}
\SetWidth{1}
\Vertex(75,0){3}
\GlueArc(75,20)(20,-90,90){4}{6}
\GlueArc(75,20)(20,90,270){4}{6}
\CCirc(75,40){10}{Black}{Blue}
\end{picture}}

\title{Ghost-gluon coupling, power corrections 
and $\Lambda_{\overline {\rm MS}}$ from twisted-mass lattice QCD at Nf=2}

\author{B. Blossier$^a$, Ph.~Boucaud$^a$, F. De soto$^b$, V.~Morenas$^c$\\
M.~Gravina$^{a}$, O. P\`ene$^a$, J.~Rodr\'iguez-Quintero$^d$ 
}

\date{}

\begin{document}
\numberwithin{equation}{section} 
\maketitle

\begin{figure}[h]
  \begin{center}
    \includegraphics{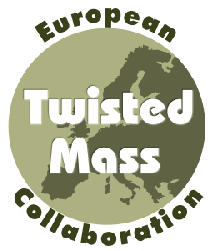}
  \end{center}
\end{figure}

\begin{center}
$^a$Laboratoire de Physique Th\'eorique\footnote{Unit\'e Mixte de Recherche 8627 du Centre National de
la Recherche Scientifique}\\
CNRS et Universit\'e  Paris-Sud XI, B\^atiment 210, 91405 Orsay Cedex,
France\\
$^b$ Dpto. Sistemas F\'isicos, Qu\'imicos y Naturales, \\
Universidad Pablo de Olavide, 41013 Sevilla, Spain.
\\
$^c$ Laboratoire de Physique Corpusculaire, Universit\'e Blaise Pascal, CNRS/IN2P3 \\
63000 Aubi\`ere Cedex, France.
\\
$^d$ Dpto. F\'isica Aplicada, Fac. Ciencias Experimentales,\\
Universidad de Huelva, 21071 Huelva, Spain.
\end{center}


\begin{abstract}

We present results concerning the non-perturbative evaluation 
of the ghost-gluon running QCD coupling constant from $N_f=2$ 
twisted-mass lattice calculations. A novel method for calibrating 
the lattice spacing, independent of the string tension and 
hadron spectrum is presented with results in agreement with previous 
estimates. The value of $\Lambda_{\overline{MS}}$ is computed 
from the running of the QCD coupling only after extrapolating to 
zero dynamical quark mass and after removing a non-perturbative 
OPE contribution that is assumed to be dominated by the 
dimension-two $\VEV{A^2}$ gluon condensate.
The effect due to the dynamical quark mass in the determination 
of $\Lams$ is discussed.

\end{abstract}

\begin{flushleft}
LPT-Orsay 10-37\\
UHU-FT/10-31
\end{flushleft}

\tableofcontents


\section{Introduction}

QCD is believed to be the theory of the  strong interactions with, as only
inputs, one mass parameter for each quark  species and the value of the QCD
coupling constant at some energy or momentum scale in some  renormalization
scheme (Alternatively, this last free parameter of the  theory can be fixed by
$\Lambda_{\rm QCD}$, the energy scale used as the typical boundary condition
for the integration of the Renormalization Group equation for the strong
coupling constant). This is the parameter which expresses the scale of strong
interactions, the only parameter in the limit of massless quarks. While the
evolution of the coupling with the momentum scale is determined by the quantum
corrections induced by the renormalization of the bare coupling and can be
computed in perturbation  theory, the strength itself of the interaction, given
at any scale by the value of the  renormalized coupling at this scale, or
equivalently by $\Lambda_{\rm QCD}$, is one of the above mentioned parameters of
the theory and has to be taken from experiment. 

The QCD running coupling can be also obtained from lattice computations, 
the free parameters  being adjusted from experimental numbers, masses, decay
constants etc. These parameters being settled, the lattice calculation of 
$\Lambda_{\rm QCD}$ proceeds in several manners:  the implemention of the
Schr\"odiger functional scheme (see,  for
instance,~\cite{Luscher:1993gh,deDivitiis:1994yp,DellaMorte:2004bc,Aoki:2009tf}
and  references therein), those based on the perturbative analysis of
short-distance sensitive lattice observables  as the ``boosted'' lattice
coupling (see for
instance~\cite{Gockeler:2005rv,Mason:2005zx,Maltman:2008bx,Davies:2008sw} and
reference therein)  and, in particular, those based on the study of  the
momentum behaviour of Green
functions(see~\cite{Alles:1996ka,Boucaud:1998bq,Boucaud:2000ey,OPEtree,OPEone,Boucaud:2002nc,Boucaud:2001qz,Sternbeck:2007br}
and references therein) are among the most extendedly applied.  Indeed, the
confrontation of  the behaviour with respect to the renormalization scale  of
2-gluon and 3-gluon Green functions with the corresponding perturbative
predictions leaves  us with a good estimate of $\alpha_S$, its running leading
to the determination  of $\Lambda_{\rm QCD}$, but also reveals a dimension-two
non-zero gluon condensate in the Landau gauge. The possible phenomenological
implications in the gauge-invariant world of such  a dimension-two gluon
condensate and in connection with confinement scenarios has been also largely 
investigated (see for instance~\cite{Gubarev:2000nz}). 

In~\cite{Boucaud:2008gn}, the Green's function approach proposed  in
ref.~\cite{Sternbeck:2007br} was followed exploiting a non-perturbative
definition of the  coupling derived from the ghost and gluon propagators for the
determination of  $\Lams$ in pure Yang-Mills ($N_f=0$). In
that work,  the renormalization scheme for the ghost-gluon vertex corresponding
to the latter coupling was properly defined. The {\it quenched} lattice results 
were analyzed over a wide momentum window, applying a ``{\it plateau}''
procedure to  extract simultaneously both $\Lams$ and the
gluon  condensate. The result is  consistent with other calculations,  for the
description of the gluon and ghost Green functions and for the running of  the
strong coupling in ``{\it quenched}'' QCD. 

In the present paper we extend~\cite{Boucaud:2008gn} to the case in which
twisted $N_f=2$ dynamical  quarks are included in the lattice
simulations for several  different bare lattice couplings
($\beta=3.9,4.05,4.2$) and different dynamical  quark masses. We use the
configurations produced by the ETM Collaboration~\cite{Boucaud:2008xu}.
This offers the opportunity  to study the effect of the quark mass on the lattice
determination of the strong  coupling, and we will see that this effect if far
from negligible. Similar works have started some time
ago~\cite{Boucaud:2001qz} using {\it unquenched}  lattice 
configurations with, at first, rather heavy  $N_f=2$ dynamical quarks, and
continuing more recently~\cite{DellaMorte:2004bc,Sternbeck:2007br} in a more
realistic case. The comparison of our current results with those needs certainly
due account of the dependence on the dynamical masses.  

Let us now summarize our strategy. For every value $\mu$ of the momentum scale
we compute, from the lattice simulations, the value of the strong coupling
constant. This can be converted via a four loops formula to a value for  
$\Lams\,(\mu)$. $\Lams$ is a scale
independent constant which sets the strong interaction scale. It results that 
$\Lams\,(\mu)$ should be independent of $\mu$ as soon as we
are in the perturbative regime. As we shall see, this is far from true at
energies of several GeV's, which are generally believed to lie in the
perturbative regime. This surprising feature was already noticed in the
quenched  case~\cite{Boucaud:2008gn}. We then need to take into account
non-perturbative contributions using Wilson expansion. In Landau gauge there
exists only one dimension two operator : $A^2\equiv A^\mu_a A_\mu^a$. The Wilson
coefficient of that operator has been computed to order
$\alpha^4$~\cite{Chetyrkin:2009kh}. We will assume that only  this  $A^2$
operator contributes~\footnote{ It is indeed not easy with present accuracy to
discard higher order operators.}. We fit $<A^2>$ so that, once the
non-perturbative contribution subtracted, one gets a good ``plateau'' for
$\Lams\,(\mu)$. We thus get an estimate of both $\Lams$ 
and $<A^2>$.

The paper is organized as follows. In section \ref{sec:alphaT}, we outline and
discuss  all the analytical tools, perturbation theory and Wilson expansion,
needed to describe the  running coupling in the appropriate renormalization
scheme. In section \ref{sec:lat},  we give the details of the lattice
computation of the coupling, describe the treatment of lattice  artefacts and
depict the analysis procedure leading to the estimate of  $\Lams$ and the gluon
condensate. The analysis is performed in section \ref{sec:anal}, where  we also
present the results and discuss  different sources of  systematical
uncertainties (special attention is paid to the higher order  contribution in
both the perturbative and the OPE expansions). Finally,  we conclude in section
\ref{sec:conclu}.

\section{The running coupling in Taylor scheme}
\label{sec:alphaT}

Among the many possibilities to the compute a strong coupling
$\alpha_S$ from lattice simulations, it has been
shown~\cite{Boucaud:2008gn} that the so-called Taylor scheme is 
among the most tractable ones because, with the help of the so-called Non-renormalization Taylor theorem, 
the coupling can be computed from two-point 
Green functions renormalized in MOM scheme. 
Following the usual notation we will write Landau gauge gluon and
ghost propagators as:
\beq\label{props}
\left( G^{(2)} \right)_{\mu \nu}^{a b}(p^2,\Lambda) &=& \frac{G(p^2,\Lambda)}{p^2} \ \delta_{a b}
\left( \delta_{\mu \nu}-\frac{p_\mu p_\nu}{p^2} \right) \ ,
\nonumber \\
\left(F^{(2)} \right)^{a,b}(p^2,\Lambda) &=& - \delta_{a b} \ \frac{F(p^2,\Lambda)}{p^2} \ ;
\eeq
with $\Lambda$ the regularisation cutoff ($a^{-1}(\beta)$ if, for instance, we specialise to lattice
regularisation). The renormalized dressing functions,
$G_R$ and $F_R$ are defined through :
\beq\label{bar}
G_R(p^2,\mu^2)\ &=& \ \lim_{\Lambda \to \infty} Z_3^{-1}(\mu^2,\Lambda) \ G(p^2,\Lambda)\nonumber\\
F_R(p^2,\mu^2)\ &=& \ \lim_{\Lambda \to \infty} \widetilde{Z}_3^{-1}(\mu^2,\Lambda)\ F(p^2,\Lambda) \ ,
\eeq
with standard MOM renormalization condition
\beq\label{bar2}
G_R(\mu^2,\mu^2)=F_R(\mu^2,\mu^2)=1 \ .
\eeq
Then, we can consider the ghost-gluon vertex which could be non-perturbatively
obtained through  a three-point Green function, defined by two ghost and one
gluon fields,  with amputated legs after dividing by two ghost and one gluon 
propagators. This vertex can be written quite generally as:
\beq\label{defGamma}
\widetilde{\Gamma}^{abc}_\nu(-q,k;q-k) = 
\ghvertex =
i g_0 f^{abc} 
\left( q_\nu H_1(q,k) + (q-k)_\nu H_2(q,k) \right) \ ,
\eeq
where $H_{1,2}$ are the two involved independent scalar form factors, $g_0$ is
the bare strong coupling,  $q$ is the outgoing ghost momentum and $k$ the
incoming one.  This bare three-pointt Green  function will be renormalized
according to:
\beq
\widetilde{\Gamma}_R=\widetilde{Z}_1 \Gamma.
\eeq
\noindent  In the MOM renormalization procedure, as explained in ref.~\cite{Boucaud:2008gn}, 
$\widetilde{Z}_1$ is fully determined by demanding 
that one specific combination of those two form factors 
(chosen at one's will) be equal to its tree-level value for a specific kinematical 
configuration; while 
\beq\label{g2R}
g_R(\mu^2) \ = \ \lim_{\Lambda \to \infty} Z_g^{-1}(\mu^2,\Lambda^2) g_0(\Lambda^2) \ = 
\ \lim_{\Lambda \to \infty}  
\frac{Z_3^{1/2}(\mu^2,\Lambda^2)\widetilde{Z}_3(\mu^2,\Lambda^2)}{ \widetilde{Z}_1(\mu^2,\Lambda^2)} 
\ g_0(\Lambda^2) \ , 
\eeq
where one puts explicitly the cut-off dependence.
If one turns now to the Taylor scheme, {\it i.e.} 
a specific MOM-type renormalization scheme defined by a kinematical configuration with 
{\bf{zero incoming ghost momentum}}, one obtains~\cite{Taylor}
\beq
\widetilde{Z}_1(\mu^2,\Lambda^2) \ \equiv 1 \ ;
\eeq
and one is left with
\beq\label{alpha} 
\alpha_T(\mu^2) \equiv \frac{g^2_T(\mu^2)}{4 \pi}=  \ \lim_{\Lambda \to \infty} 
\frac{g_0^2(\Lambda^2)}{4 \pi} G(\mu^2,\Lambda^2) F^{2}(\mu^2,\Lambda^2) \ ;
\eeq
where we also apply the renormalization condition for the propagators, eqs. (\ref{bar},\ref{bar2}), 
to replace the renormalization constants, $Z_3$ and $\widetilde{Z}_3$, by the bare dressing 
functions. As emphasized in ~\cite{Boucaud:2008gn}, the remarkable feature of  
\eq{alpha} is that  it involves only $F$ and $G$ so that no measure of the ghost-gluon vertex 
is needed for the determination of the coupling constant.

\subsection{Pure perturbation theory}
\label{PTh}

The Taylor coupling and the one renormalized in the standard $\overline{\rm MS}$ prescription, as any other 
different definitions of the coupling constant, can be related through:
\beq\label{alpha2}
\alpha_T(\mu^2) \ = \ \overline{\alpha}(\mu^2) \ \left( 1 + \sum_{i=1} c_i 
\left( \frac{\overline{\alpha}(\mu^2)}{4 \pi}\right)^i \ \right) \ ;
\eeq
where the coefficients $c_i$'s can be obtained in perturbation 
theory~\cite{Chetyrkin00,Chetyrkin:2004mf}
\beq\label{cis}
c_1 &=& \frac{507-40 N_f}{36} \ ,
\nonumber \\
c_2 &=& \frac{76063}{144} - \frac{351} 8 \zeta(3) - 
    \left( \frac{1913}{27} + \frac 4 3 \zeta(3) \right) \ N_f 
    + \frac{100}{91} N_f^2 
\nonumber \\
c_3 &=& \frac{42074947}{1728} - \frac{60675}{16}\zeta(3) 
- \frac{70245}{64} \zeta(5) - \left( \frac{769387}{162} - 
\frac{8362}{27} \zeta(3) -\frac{2320}{9} \zeta(5) \right) \ N_f 
\nonumber \\
&+& \left( \frac{199903}{972} + \frac{28}{9} \zeta(3) \right) \ N_f^2
- \frac{1000}{729} \ N_f^3
\ .
\eeq
It was proven in ref.~\cite{Boucaud:2008gn} that these coefficients could 
be also directly derived from the anomalous dimensions for gluon and ghost propagators, 
as \eq{alpha} indicates. 
The three coefficients in \eq{cis} obviously define unambigously the running of $\alpha_T$ up to 
four-loops given by
\begin{align}
  \label{betainvert}
  \begin{split}
      \alpha_T(\mu^2) &= \frac{4 \pi}{\beta_{0}t}
      \left(1 - \frac{\beta_{1}}{\beta_{0}^{2}}\frac{\log(t)}{t}
     + \frac{\beta_{1}^{2}}{\beta_{0}^{4}}
       \frac{1}{t^{2}}\left(\left(\log(t)-\frac{1}{2}\right)^{2}
     + \frac{\widetilde{\beta}_{2}\beta_{0}}{\beta_{1}^{2}}-\frac{5}{4}\right)\right) \\
     &+ \frac{1}{(\beta_{0}t)^{4}}
 \left(\frac{\widetilde{\beta}_{3}}{2\beta_{0}}+
   \frac{1}{2}\left(\frac{\beta_{1}}{\beta_{0}}\right)^{3}
   \left(-2\log^{3}(t)+5\log^{2}(t)+
\left(4-6\frac{\widetilde{\beta}_{2}\beta_{0}}{\beta_{1}^{2}}\right)\log(t)-1\right)\right)
   \end{split}
\end{align}
with $t=\ln \frac{\mu^2}{\Lambda_T^2}$, since the coefficients of the $\beta$-function of $\alpha_T$, 
\beq\label{beta}
\beta_T(\alpha_T) \ = \ 
\frac{d\alpha_T}{d\ln{\mu^2}} \ = \ - 4 \pi \ 
\sum_{i=0} \widetilde{\beta}_i \left( \frac{\alpha_T} {4 \pi} \right)^{i+2} \ ,
\eeq
can be derived, 
\beq\label{betacoefs}
\widetilde{\beta}_0 &=& \overline{\beta}_0 = 11 - \frac 2 3 N_f
\nonumber \\
\widetilde{\beta}_1 &=& \overline{\beta}_1 = 102 - \frac{38} 3 N_f
\nonumber \\
\widetilde{\beta}_2 &=& \overline{\beta}_2 -\overline{\beta}_1 c_1 + \overline{\beta}_0 (c_2-c_1^2)
\nonumber \\
&=&3040.48 \ - \ 625.387 \ N_f \ + \ 19.3833 \ N_f^2
\nonumber \\
\widetilde{\beta}_3 &=& \overline{\beta}_3 - 2 \overline{\beta}_2 c_1 + \overline{\beta}_1 c_1^2
+ \overline{\beta}_0 (2 \ c_3 - 6 \ c_2 c_1 + 4 \ c_1^3)
\nonumber \\
&=&  100541 \ - \ 24423.3 \ N_f \ + \ 1625.4 \ N_f^2 \ - \ 27.493 \ N_f^3
\ ,
\eeq
from the knowledge of those coefficients, $c_i$'s, and from that of the standard 
$\overline{\rm MS}$ $\beta$-function, 
\beq\label{betaMS}
\beta_{\overline{\rm MS}}(\overline{\alpha}) \ = \ 
\frac{d\overline{\alpha}}{d\ln{\mu^2}} \ = \ - 4 \pi \ 
\sum_{i=0} \overline{\beta}_i \left( \frac{\overline{\alpha}} {4 \pi} \right)^{i+2} 
\eeq
given at four loops in ref.~\cite{Larin} ($\overline{\beta}_0$ and
$\overline{\beta}_1$  being scheme-independent).  As for  the $\Lambda_{\rm
QCD}$ parameters in the two schemes, they are  related through
\beq\label{ratTMS}
\frac{\Lambda_{\overline{\rm MS}}}{\Lambda_T} \ = \ e^{\displaystyle -\frac{c_1}{2 \beta_0}} \ = \ 
e^{\displaystyle - \frac{507-40 N_f}{792 - 48 N_f}}\ = \ 0.541449
\ .
\eeq
Then, the lattice data for the coupling, obtained through \eq{alpha}, 
can be confronted to the perturbative formulae, \eq{betainvert} with the 
$\beta$-function coefficients given by \eq{cis}, over the large-momentum 
window where the four-loop perturbation theory is reliable.

\subsection{OPE power corrections}
\label{OPEsection}

As was previously done for the quenched analysis in \cite{Boucaud:2008gn},  in
order to take full advantage of the lattice data (and reduce the systematic
uncertainties) when confronting them with a formula for the QCD running
coupling, one needs to take into account the  gauge-dependent OPE
power corrections (cf.~\cite{OPEtree,OPEone,Boucaud:2002nc})  to  $\alpha_T$.
In Landau gauge there exists only one dimension-two operator allowed to have a
vacuum expectation value: $A^2 \equiv A^2\equiv A^\mu_a A_\mu^a$. We will stick
to it, leaving the discussion of higher dimension operators to 
section~\ref{sec:anal}. We will also not go beyond the leading logs in this
section, leaving again the discussion of higher orders~\cite{Chetyrkin:2009kh} 
to section~\ref{sec:anal}. 

The leading power contribution to the ghost and gluon propagators 
can thus be computed using the operator product 
expansion~\cite{Wilson69} (OPE), as is done in ref.~\cite{Boucaud:2005xn},
and one obtains:
\beq\label{OPE1}
(F^{(2)})^{a b}(q^2) &=& (F_{\rm pert}^{(2)})^{a b}(q^2) \ + \ 
w^{a b} \ \frac{\langle A^2 \rangle}{4 (N_C^2-1)} \ + \ \dots 
\nonumber \\
(G^{(2)})_{\mu\nu}^{a b}(q^2) &=& (G_{\rm pert}^{(2)})_{\mu\nu}^{a b}(q^2) \ + \ 
w_{\mu\nu}^{a b} \ \frac{\langle A^2 \rangle}{4 (N_C^2-1)} \ + \ \dots 
\eeq
where the Wilson coefficients, diagramatically expressed as follows
(the bubble means contracting the 
color and Lorentz indices of the incoming legs with 
$1/2 \delta_{st} \delta_{\sigma \tau}$) 
\beq\label{OPE3}
w^{a b} \ &=&   2 \times \rule[0cm]{0cm}{1.7cm} \ghost, 
\nonumber \\
w_{\mu\nu}^{a b} \ &=&  \gluonTwoB
+ \
 2 \times \rule[0cm]{0cm}{1.7cm} \gluonTwoA \ ,
\eeq
can be computed by invoking the SVZ factorisation~\cite{SVZ}. 
Thus, after some algebra and the appropriate renormalization at the subtraction point $q^2=\mu^2$, 
according to the MOM scheme definition (details of the computation can be found in 
\cite{Boucaud:2008gn,Boucaud:2005xn,OPEone}), one gets at tree level
\beq\label{Z3fantome}
F_R(q^2,\mu^2) \ = \ F_{R, {\rm pert}}(q^2,\mu^2) \
\left(  1 + \frac{3}{q^2} \frac{g^2_R \langle A^2 \rangle_{R,\mu^2}} {4 (N_C^2-1)} \right) 
\nonumber \\ 
G_R(q^2,\mu^2) \ = \ G_{R, {\rm pert}}(q^2,\mu^2) \
\left(  1 + \frac{3}{q^2} \frac{g^2_R \langle A^2 \rangle_{R,\mu^2}} {4 (N_C^2-1)} \right) \ ,
\eeq
where the multiplicative correction to the purely perturbative contributions
were determined  up to corrections of the order $1/q^4$ or $\ln{q/\mu}$. 
Finally, putting together the defining relation \eq{alpha} and the results
eqs.~(\ref{Z3fantome})  we obtain
\beq\label{alphahNPt}
\alpha_T(\mu^2) &=& \lim_{\Lambda \to \infty} 
\frac{g_0^2}{4 \pi} F^2(\mu^2,\Lambda) G(\mu^2,\Lambda) \nonumber \\
&=& 
\lim_{\Lambda \to \infty} 
\frac{g_0^2}{4 \pi} F^2(q_0^2,\Lambda) F^2_R(\mu^2,q_0^2) \
G(q_0^2,\Lambda)  G_R(\mu^2,q_0^2) 
\nonumber \\
&=& \alpha^{\rm pert}_T(q_0^2) \  
F^2_{R, {\rm pert}}(\mu^2,q_0^2) \ G_{R, {\rm pert}}(\mu^2,q_0^2) 
\ \left(  1 + \frac{9}{\mu^2} \frac{g^2_T(q_0^2) \langle A^2 \rangle_{R,q_0^2}} {4 (N_C^2-1)}
\right) \nonumber \\
&=& 
\alpha^{\rm pert}_T(\mu^2)
\ \left(  1 + \frac{9}{\mu^2} \frac{g^2_T(q_0^2) \langle A^2 \rangle_{R,q_0^2}} {4 (N_C^2-1)}
\right) \ ,
\eeq
where $q_0^2 \gg \Lambda^2_{\rm QCD}$ is some perturbative scale and the
$\beta$-function, and its coefficients in \eq{betacoefs}, of course describe the
running  of the perturbative part of the evolution, $\alpha_T^{\rm pert}$. The
anomalous dimension for the  Wilson coefficient,
\beq\label{gA2}
\gamma_{A^2}(\alpha(\mu^2)) \ = \ 
\lim_{\Lambda \to \infty} \ \frac{d}{d\ln\mu^2} \ln Z_{A^2}(\mu^2,\Lambda^2) \ 
= \ - \gamma_0^{A^2} \ \frac{\alpha(\mu^2)}{4 \pi} \ + \dots 
\eeq
where $A_R^2 = Z_{A^2}^{-1} A^2$, is neglected in \eq{alphahNPt}. 

The  leading logarithm contribution for the Wilson coefficient are
incorporated as explained in~\cite{Boucaud:2008gn}, yielding:
\beq\label{alphahNP}
\alpha_T(\mu^2)
\ = \
\alpha^{\rm pert}_T(\mu^2)
\ 
\left( 
 1 + \frac{9}{\mu^2} 
\left( 
\frac{\alpha^{\rm pert}_T(\mu^2)}{\alpha^{\rm pert}_T(q_0^2)}
\right)^{1-\gamma_0^{A^2}/\beta_0}
\frac{g^2_T(q_0^2) \langle A^2 \rangle_{R,q_0^2}} {4 (N_C^2-1)}
\right) \ ,
\eeq
where $\gamma_0^{A^2}$ can be taken from \cite{Gracey:2002yt,Chetyrkin:2009kh} to give
\beq\label{gA0}
1 - \frac{\gamma_0^{A^2}}{\beta_0} 
\ = \ 1 - \frac{105 - 8 N_f}{132 - 8 N_f}
\ = \ \frac{9}{44 - \frac{8}{3} N_f} \ ,
\eeq
which agrees for $N_f=0$ with the power of the logarithmic correction applied, 
and shown to have a negligible impact on $\alpha$ in~\cite{Boucaud:2008gn}. We shall first 
apply formula~\eq{alphahNP}, approximated up to the four-loop level in 
perturbation and up to the leading-log in the OPE expansion, to describe the 
lattice data in the next sections. As already mentioned we postpone the use
of four-loop Wilson coefficients~\cite{Chetyrkin:2009kh} and the study of the
impact of higher order operators to the end of section~\ref{sec:anal}.

\section{The lattice computation of the Taylor coupling}
\label{sec:lat}

The following section is devoted to the computation of the running coupling in 
Taylor scheme, \eq{alpha}, from the lattice. 
The results presented here are based on the gauge field 
configurations generated by the European Twisted Mass Collaboration 
(ETMC) with the tree-level improved Symanzik gauge action~\cite{Weisz:1982zw}  
and the twisted mass fermionic action~\cite{Frezzotti:2000nk} at
maximal twist. 

\subsection{The lattice action}

A very detailed discussion about the twisted mass and tree-level improved
Symanzik  gauge actions, and about the way they are implemented by ETMC, can be
found in 
refs.~\cite{Boucaud:2007uk,Boucaud:2008xu,Urbach:2007rt,Dimopoulos:2008sy}. 
Here, for the sake of completeness, we will present a brief reminder of the
twisted action and the run parameters for the gauge configurations that will be
exploited in the present work (See tab.~\ref{setup}).

The Wilson twisted mass fermionic lattice action for two flavours of mass
degenerate  quarks, reads  (in the so called twisted
basis~\cite{Frezzotti:2000nk,Frezzotti:2003ni} ) 
\begin{align}
  \label{eq:Sf} 
  \begin{split} 
    S_\mathrm{tm}^{\rm F} = &\, a^4\sum_x\Bigl\{ 
    \bar\chi_x\left[D_{\rm W}+ m_0 + i\gamma_5\tau_3\mu_q  
    \right]\chi_x\Bigr\}\, , \\ 
    & D_{\rm W} = \frac{1}{2}\gamma_\mu\left(\nabla_\mu+\nabla_\mu^{*}\right) 
    -\frac{ar}{2}\nabla_\mu\nabla_\mu^{*} \, ,
  \end{split} 
\end{align}
where $m_0$ is the bare untwisted quark mass and $\mu_q$ the bare twisted 
quark mass, $\tau_3$ is the third Pauli matrix acting in flavour space  
and $r$ is the Wilson parameter, which is set to $r=1$ in the simulations.
The operators $\nabla_\mu$ and $\nabla_\mu^{*}$ stand for the gauge covariant nearest  
neighbour forward and backward lattice derivatives. 
The bare quark mass $m_0$ is related as usual to the so-called hopping  
parameter $\kappa$, by $\kappa=1/(8+2am_0)$. Twisted mass fermions are said to be at  
{\em maximal twist} if the bare untwisted mass is tuned to its critical  
value, $m_\mathrm{crit}$. This is in practice done by setting the so-called untwisted
PCAC mass to zero.

In the gauge sector  the  tree-level Symanzik improved 
gauge action (tlSym)~\cite{Weisz:1982zw} is applied. This action includes besides the 
plaquette term $U^{1\times1}_{x,\mu,\nu}$ also rectangular $(1\times2)$ Wilson loops 
$U^{1\times2}_{x,\mu,\nu}$. It reads  
\beq 
  \label{eq:Sg} 
    S_g =  \frac{\beta}{3}\sum_x\Biggl(  b_0\sum_{\substack{ 
      \mu,\nu=1\\1\leq\mu<\nu}}^4\{1-\re\tr(U^{1\times1}_{x,\mu,\nu})\}\Bigr.  
     \Bigl.+ 
    b_1\sum_{\substack{\mu,\nu=1\\\mu\neq\nu}}^4\{1 
    -\re\tr(U^{1\times2}_{x,\mu,\nu})\}\Biggr)\, , 
\eeq
where $\beta \equiv 6 / g_0^2$, $g_0$ being the bare lattice  coupling and it
is set    $b_1=-1/12$ (with $b_0=1-8b_1$ as dictated by the requirement  of
continuum limit normalization). Note that at $b_1=0$ this action becomes the
usual Wilson plaquette gauge action.  The run parameters for $\beta$ and $\mu_q$
of the gauge configurations that will  be exploited in the following can be
found in tab.~\ref{setup}.

\begin{table}[ht]
\centering
\begin{tabular}{||c|c|c|c||}
\hline
\hline
$\beta$ & $a \mu_q$ & Volume & Number of confs.
\\ \hline
$3.9$ &  
\begin{tabular}{c}
0.004 \\
0.0064 \\
0.010
\end{tabular}
& 
$24^3\times48$ & 
\begin{tabular}{c}
$120$ \\
$20$ \\
$20$
\end{tabular}
\\ \hline
$4.05$ &  
\begin{tabular}{c}
0.003 \\
0.006 \\
0.008 \\
0.012
\end{tabular}
& $32^3\times64$ & 
\begin{tabular}{c}
$20$ \\
$20$ \\
$20$ \\
$20$
\end{tabular}
\\ \hline
$4.2$ & 0.0065 & $32^3\times64$ & $200$
\\ \hline
\hline
\end{tabular}
\caption{Run parameters of the exploited data from ETMC collaboration.}
\label{setup}
\end{table}

\subsection{The computation of the gluon and ghost Green functions}

Applying \eq{alpha} demands to compute the gauge-fixed 2-point gluon and ghost 
Green functions from the lattice. To this goal, we exploited ETMC gauge
configurations obtained for  $\beta=3.9$, $\beta=4.05$ and $\beta=4.2$ and a
large variety of dynamical quark masses, fixed by  the values of the $\mu_q$
parameter. The lattice gauge configurations are transformed to Landau gauge
by minimising the following functional of the SU(3) matrices, $U_\mu(x)$,
\beq
F_U[g] = \mbox{\rm Re}\left[ \sum_x \sum_\mu  \hbox{Tr}\left(1-\frac{1}{N}g(x)U_\mu(x)g^\dagger(x+\mu) \right) \right]
\eeq
with respect to the gauge transform $g$, by applying a combination of
overrelaxation algorithm and  Fourier acceleration~\footnote{We end when 
$|\partial_\mu A_\mu|^2 <10^{-11}$ and when the spatial integral of $A_0$ is
constant in time to better than $10^{-6}$.}. This procedure does not avoid  the
possibility of lattice Gribov copies that, in any case, have  been reported to
have a nonsignificant influence beyond the lowest momenta. Then, the gauge field
is defined as
\beq 
A_\mu(x+ \hat \mu/2) = \frac {U_\mu(x) - U_\mu^\dagger(x)}{2 i a g_0}
- \frac 1 3 \hbox{Tr}\left (\frac{U_\mu(x) - U_\mu^\dagger(x)}
{2 i a g_0}\right ) \label{amu}
\eeq
where $\hat \mu$ indicates the unit lattice vector in the direction $\mu$ and $g_0$ is
the bare coupling constant. The 2-gluon Green functions is computed in momentum space by
\beq
 \left( G^{(2)}\right)^{a_1 a_2}_{\mu_1\mu_2}(p)=
 \langle A_{\mu_1}^{a_1}(p)A_{\mu_2}^{a_2}(-p) \rangle
\label{greenG}
\eeq
where $\langle \cdots \rangle$ indicates the Monte-Carlo average
and where
\beq
	A_\mu^a(p)=\frac 1 2 \hbox{Tr}\left [\sum_x A_\mu(x+ \hat \mu/2)
	 \exp(i p (x+ \hat \mu/2))\lambda^a\right]\label{amufour}
\eeq
$\lambda^a$ being the Gell-Mann matrices and the trace being taken in the
$3\times 3$ color space. 

On the other hand, the ghost propagator is also computed in Landau gauge,
\beq\label{greenF}
\left( F^{(2)} \right)^{ab}(x-y) \ \equiv \ \langle \left( M^{-1} \right)^{ab}_{xy} \rangle \ ,
\eeq
as the inverse of 
the Faddeev-Popov operator, that is written as the lattice divergence,
\beq
M(U) = -\frac{1}{N} \nabla \cdot \widetilde{D}(U)
\eeq
where the operator $\widetilde{D}$ acting on an arbitrary 
element of the Lie algebra, $\eta$ reads:
\beq
\widetilde{D}(U) \eta(x) = \frac{1}{2} \left(U_{\mu}(x)\eta(x+\mu) -
\eta(x)U_\mu(x)+\eta(x+\mu)U_\mu^\dagger-U_\mu^\dagger(x)\eta(x)\right)
\ . 
\eeq
More details on the lattice procedure for the inversion of 
Faddeev-Popov operator can be found in~\cite{Boucaud:2005gg}.

\subsection{On the treatement of lattice artefacts}

The lattice estimates of the Landau-gauge propagators through
eqs.~(\ref{greenG},\ref{greenF}),  after Fourier transforming the ghost
correlator and appropriate projection of both  as indicated by \eq{props}, lead
to the determination of ghost and gluon dressing functions to be used in 
\eq{alpha}. Both dressing functions are dimensionless lattice correlation
functions (let us note  both as $Q \equiv F,G$) that, because of general
dimensional arguments, depend on the lattice  momentum $a\,{p}_\mu$,
%
%
where 
\beq\label{pmu}
p_\mu = \frac{2\pi n}{N a} \qquad n=0,1,\cdots,N \ ,
\eeq
 and on the strong interaction scale $\Lambda_{\rm QCD}$. Anticipating on the averaging 
 over hypercubic orbits and on the treatment 
 of hypercubic lattice artefacts we get  
 $Q \equiv Q(a^2\,{p}^2,a^2\Lambda_{\rm QCD}^2)$.
Our choice for the lattice action ensures that the discretization artefacts 
due to the lattice are $\mathcal{O}(a^2)$, where $a$ is the lattice spacing. 

Thus, the running coupling in Taylor scheme is obtained by
\beq\label{alphaT}
\alpha_T(\mu^2) \ = \ \lim_{a \to 0} \frac {g(a^2)}{4 \pi} F^2\left(a^2\,{p}^2,a^2\Lambda_{\rm QCD}^2\right) 
G\left(a^2\,{p}^2,a^2\Lambda_{\rm QCD}^2\right) \ ,
\eeq
where taking the limit of a vanishing lattice spacing indeed implies the proper elimination
of lattice artefacts.

\subsubsection{Hypercubic $H(4)$-extrapolation}
     
A first kind of artefacts that can be systematically 
cured~\cite{Becirevic:1999uc,deSoto:2007ht} are those due to the
breaking of the  rotational symmetry of the Euclidean space-time when using an
hypercubic lattice,  where this symmetry is restricted to the discrete $H(4)$
isometry group. It is convenient to compute first the average  of any
dimensionless lattice quantity $Q(a p_\mu)$ over every orbit of the group
$H(4)$. In general several orbits of $H(4)$ correspond to one value of $p^2$.
Defining the $H(4)$ invariants
 \beq
 p^{[4]}=\sum_{\mu=1}^{4} p_\mu^4\qquad p^{[6]}=\sum_{\mu=1}^{6} p_\mu^6
 \eeq
it happens that the orbits of $H(4)$ are 
labelled~\footnote{On totally general grounds, any $H(4)$-invariant 
polynome can be written only in terms of the four 
invariants $p^{[2 i]}$ with $i=1,2,3,4$~\cite{Becirevic:1999uc,deSoto:2007ht}. 
As a consequence of the upper cut for momenta, the first three of these invariants suffice 
to label all the orbits we deal with and hence any presumed dependence on $p^{[8]}$ is  
neglected.} 
by the set $p^2, a^2 p^{[4]},
a^4 p^{[6]}$. In the continuum limit the effect of $ a^2 p^{[4]},
a^4 p^{[6]}$ vanishes. We can thus define the quantity $Q(a p_\mu)$ averaged over 
$H(4)$ as
\beq\label{Q246}
Q(a^2\,p^2, a^4p^{[4]}, a^6 p^{[6]}, a^2\Lambda_{\rm QCD}^2).
\eeq

If the lattice spacing is small enough such that $\epsilon=a^2 p^{[4]}/p^2 \ll 1$,
the dimensionless lattice correlation function defined in~\eq{Q246} can be
expanded in powers of $\epsilon$:
\beq
Q(a^2\,{p}^2, a^4p^{[4]}, a^6 p^{[6]},a^2\Lambda_{\rm QCD}^2)
= Q(a^2p^2,a^2\Lambda_{\rm QCD}^2) +
\left.\frac{dQ}{d\epsilon}\right|_{\epsilon=0} a^2
\frac{p^{[4]}}{p^2} + \cdots
\eeq
$H(4)$ methods are based on the appearance of a $\mathcal{O}(a^2)$
corrections driven by a $p^{[4]}$ term. The basic method is to fit
from the whole set of orbits sharing the same $p^2$ the
coefficient $dQ/d\epsilon$ and get the extrapolated value of $Q$,
free from $H(4)$ artefacts. If we further assume that the coefficient 
\[
\left.R(a^2p^2,a^2\Lambda_{\rm QCD}^2) = 
\frac{dQ\left(a^2p^2 ,0,0,a^2\Lambda_{\rm QCD}^2\right)}{d\epsilon}\right|_{\epsilon=0}
\]
has a smooth dependence on $a^2p^2$ over a given momentum window, we can 
expand $R$ as $R=R_0+R_1 a^2p^2$ and make a global fit in a
momentum window between $(p-\delta,p+\delta)$ to extract the extrapolated value
of $Q$ for the momenta $p$ in the window, and shift to the next window  etc. 
This procedure of fitting with windows is somehow different from the 
basic one,  since the extrapolation does not rely on any  particular
assumption for the functional form of $R$. On the other, the systematic error
coming from the extrapolation can be estimated by modifying the width of the
fitting window.

It is worthwile to mention that we considered in this work  anisotropic lattice
of the type $L^3$x$T$, with $T=2L$.  This finite volume effect reduces the $H(4)$
lattice symmetry to $H(3)$.  Deviations from $H(4)$  are to be expected in the
long-distance physics. But ultraviolet physics should not be affected.  As far
as we are interested in the high-momentum regime,  we will assume the previous
treatement of the lattice artefacts to be valid.

\subsubsection{Quark mass artefacts}

In this section we will consider the influence of the dynamical quark masses. 
We will argue that this is a ${\mathcal O(a^2 \mu_q^2)}$ effect and therefore 
that it is a lattice artefact.

Following the treatement described in the previous section, we have calculated 
the $H(4)$-free ghost and gluon dressing functions, that we shall denote in the following 
by $\widehat{F},\widehat{G}$. These dressing functions can be combined in order to calculate 
the $H(4)$-free lattice coupling through~\eq{alphaT}.

In the analysis performed ref.~\cite{Boucaud:2008gn} by exploiting ``quenched'' 
configurations, a pretty good scaling of $\alpha_T$ was found, computed at different 
 $\beta$'s from \eq{alphaT}, once one parameter describing the lattice
 spacing ratios had been fitted. This seems to imply that the residual 
$\mathcal{O}(a^2 p^2)$ artefacts are negligible for the ``quenched'' coupling
constant in Taylor scheme after $H(4)$-extrapolation has been performed.  
Of course $\mathcal{O}(a^2 \Lambda_{\rm QCD}^2)$ artefacts may still be hidden
in the matching coefficient between lattice spacings.

 It is worth pointing that all the divergent contributions
appearing for  the dressing functions or the bare coupling when the lattice
spacing vanishes cancel when combined in \eq{alphaT}.

\begin{figure}[hbt]
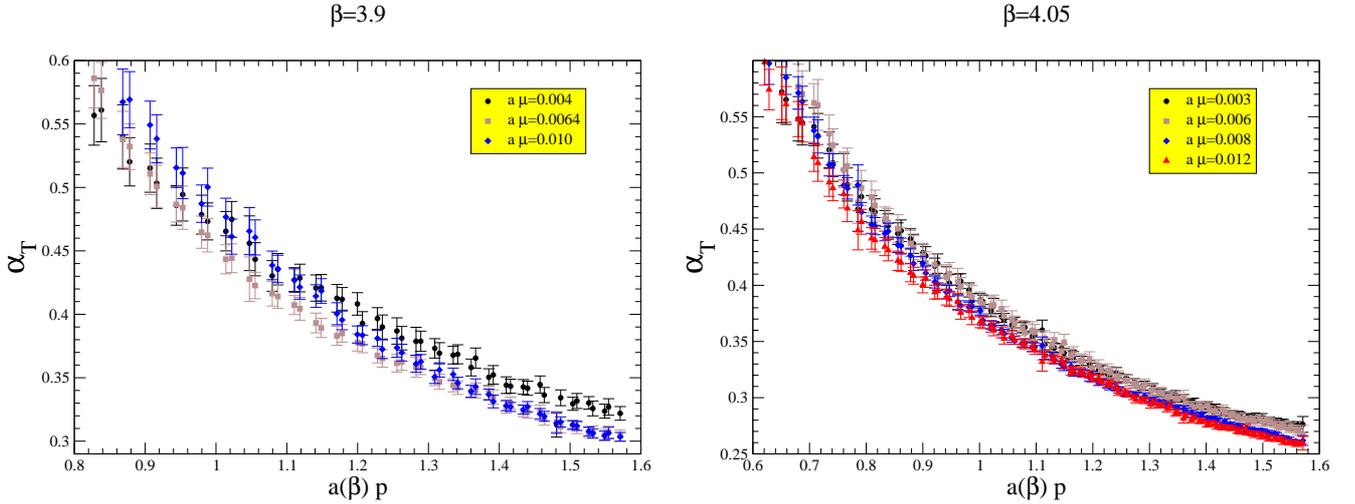

\begin{center}
\begin{tabular}{cc}
\includegraphics[width=8.5cm]{figs/aMOMb39.eps}
&
\includegraphics[width=8.5cm]{figs/aMOMb405.eps}
\end{tabular}
\end{center}
\caption{\small The Taylor couplings estimates through \eq{alphaT}, after $H(4)$-extrapolation, 
at $\beta=3.9$ for $\mu_q=0.004,0.0064,0.010$ (left-hand plots) and at $\beta=4.05$ 
for $\mu_q=0.003,0.006,0.008,0.012$ (right-hand plots).}
\label{fig:brut}
\end{figure}

However, when analyzing ``unquenched'' lattice configurations, one should keep
in mind that  one additional mass scale, the dynamical quark mass, is playing
a role. In Fig.~\ref{fig:brut} one can see the Taylor coupling after  hypercubic
extrapolation for different $\mu_q$ at fixed $\beta=3.9$ and $4.05$. Indeed,
a dependence in $\mu_q$ is clearly seen. If it is an artefact the dependence
should be in $a^2 \mu_q^2$. If it is an effect in the continuum it should be
some unknown function of the physical mass $\mu_q$. 
Trying an $\mathcal{O}(a^2 \mu_q^2)$ dependence, we write the expansion :
\beq
\widehat{\alpha}_{T}(a^2p^2,a^2\mu_q^2) &=& \frac{g_0^2(a^2)}{4 \pi}
\widehat{G}(a^2p^2,a^2\mu_q^2) \widehat{F}^{2}(a^2 p^2,a^2\mu_q^2) \nonumber \\ 
&=& \widehat{\alpha}_{T}(a^2p^2,0) + 
\frac{\partial \widehat{\alpha}_{T}}{\partial (a^2\mu_q^2)} \left(a^2 p^2 \right) \ 
a^2 \mu_q^2 \ + \ \cdots ; 
\label{H4exp}
\eeq
%

\begin{figure}[hbt]
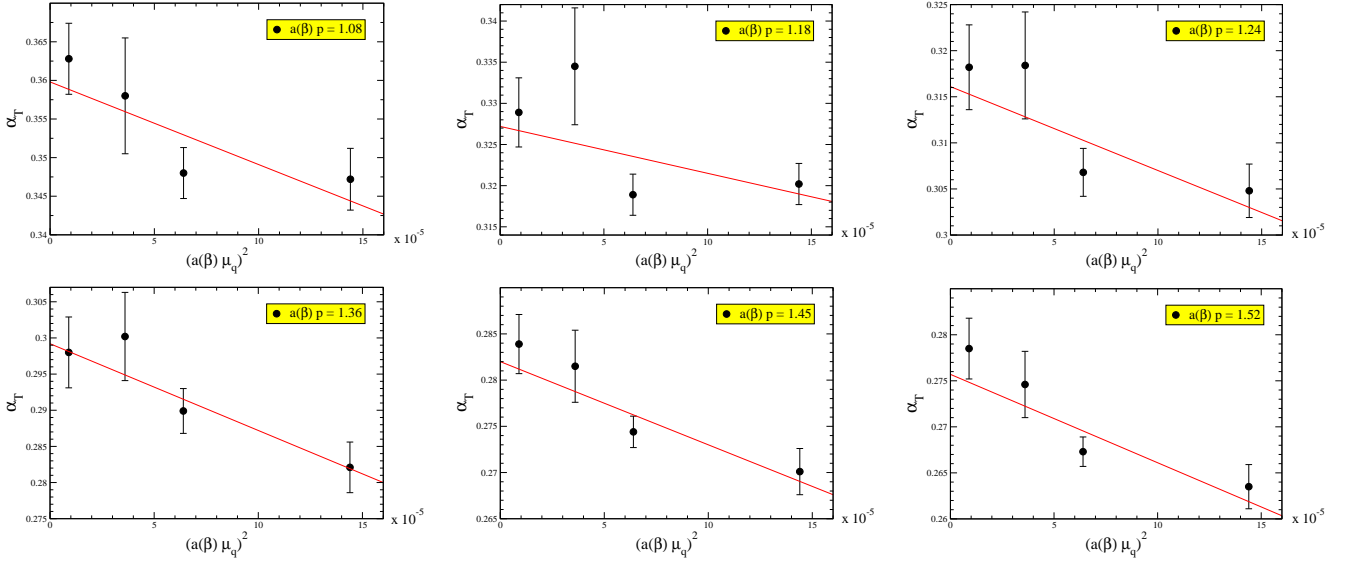

\begin{center}
\begin{tabular}{ccc}
\includegraphics[width=5.5cm]{figs/slope108.eps} 
& 
\includegraphics[width=5.5cm]{figs/slope118.eps}
&
\includegraphics[width=5.5cm]{figs/slope124.eps} 
\\ 
\includegraphics[width=5.5cm]{figs/slope136.eps}
&
\includegraphics[width=5.5cm]{figs/slope145.eps} 
& 
\includegraphics[width=5.5cm]{figs/slope152.eps}
\end{tabular}
\end{center}
\caption{\small We plot the values of the Taylor coupling at $\beta=4.05$, computed 
for some representative values of the lattice momentum, $a(4.05) p=1.08,1.18,1.24,1.36,1.45,1.52$, 
in terms of $a^2(4.05) \mu_q^2$ and show the suggested linear extrapolation at $a^2\mu_q^2=0$.}
\label{fig:slopes}
\end{figure}

\begin{figure}[hbt]
\begin{center}
\includegraphics[width=15cm,height=9cm]{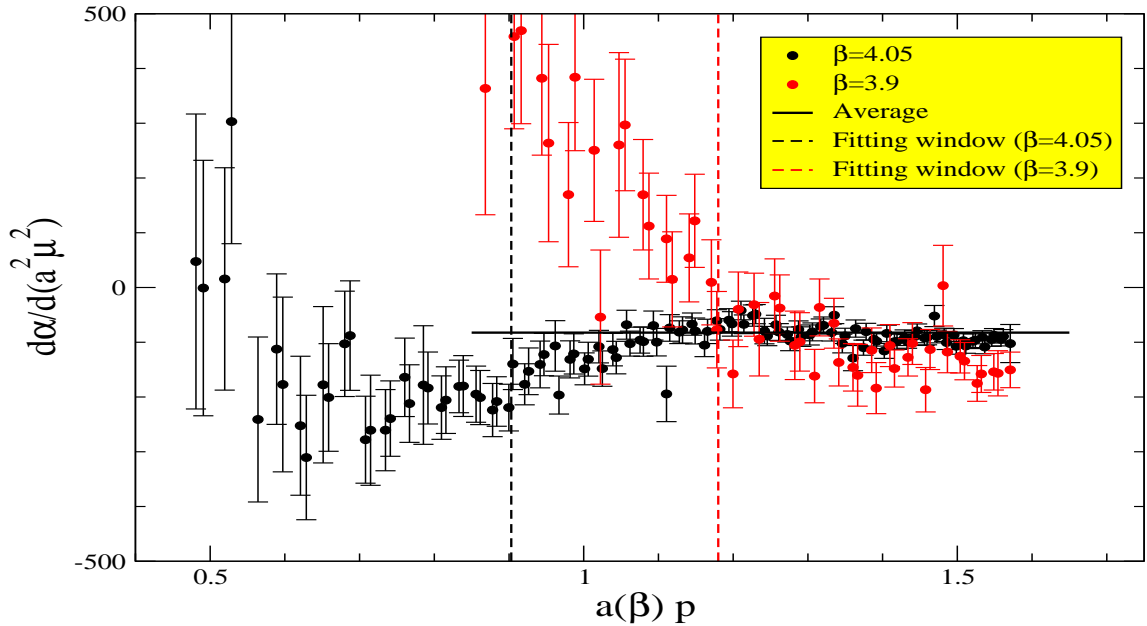} 
\end{center}
\caption{\small The slopes for the mass squared extrapolation in terms of $a p$
computed for the four lattices simulations at $\beta=4.05$ ($32^3\times 64$) with $a \mu_q=0.003,0.006,0.008,0.012$ and 
for the three ones at $\beta=3.9$ ($24^3\times 48$) with $a \mu_q=0.004,0.0064,0.010$. The supposed constant behaviour 
appear to be reached when lattice volume effects become negligible for each simulation.}
\label{fig:C0}
\end{figure}

Provided that the first-order expansion in \eq{H4exp} is reliable,  a linear
behaviour on $a^2 \mu_q^2$ has to be expected for the  lattice estimates of
$\widehat{\alpha}_T$ for any fixed lattice  momentum computed from simulations
at any given $\beta$ and several   values of $\mu_q$. We explicitely check this
linear behaviour to occur  for the results from our $\beta=4.05$ and $\beta=3.9$
simulations and  show in Fig.~\ref{fig:slopes} some plots of
$\widehat{\alpha}_T$ computed at $\beta=4.05$  (where four different quark masses
are available) for some representatives lattice momenta  in terms of
$a^2\mu_q^2$. We thus write the Taylor expansion as
and after neglecting the $\mathcal{O}(a^4)$ contributions get
\beq
\widehat{\alpha}_T(a^2p^2,a^2\mu_q^2) 
= \alpha_T(p^2) + R_0(a^2p^2) \ a^2 \mu_q^2 ,
\label{alpha:mu} 
\eeq
where $R_0(a^2p^2)$ is defined as 
\beq
R_0(a^2 p^2) \equiv
\frac{\partial \widehat{\alpha}_{T}}{\partial (a^2\mu_q^2)} \ 
\eeq
In fig.~\ref{fig:C0}, we plot $R_0(a^2p^2)$ as a function of
 $a p$ computed for the four lattices simulations at $\beta=4.05$ with
different quark masses and  for the three ones at $\beta=3.9$ (see
tab.~\ref{setup}).  Indeed, it can be seen that a constant behaviour appears to 
be achieved for $ p \ge p_{\rm min} \simeq 2.8$ GeV. We will not risk an
interpretation of the data below $(a p)_{\rm min}$. The striking observation
here is that above $p_{\rm min}$ both lattice spacings exhibit a fairly constant
 $R_0(a^2p^2)$ and a good enough scaling between both $\beta$'s.

\begin{figure}[hbt!]
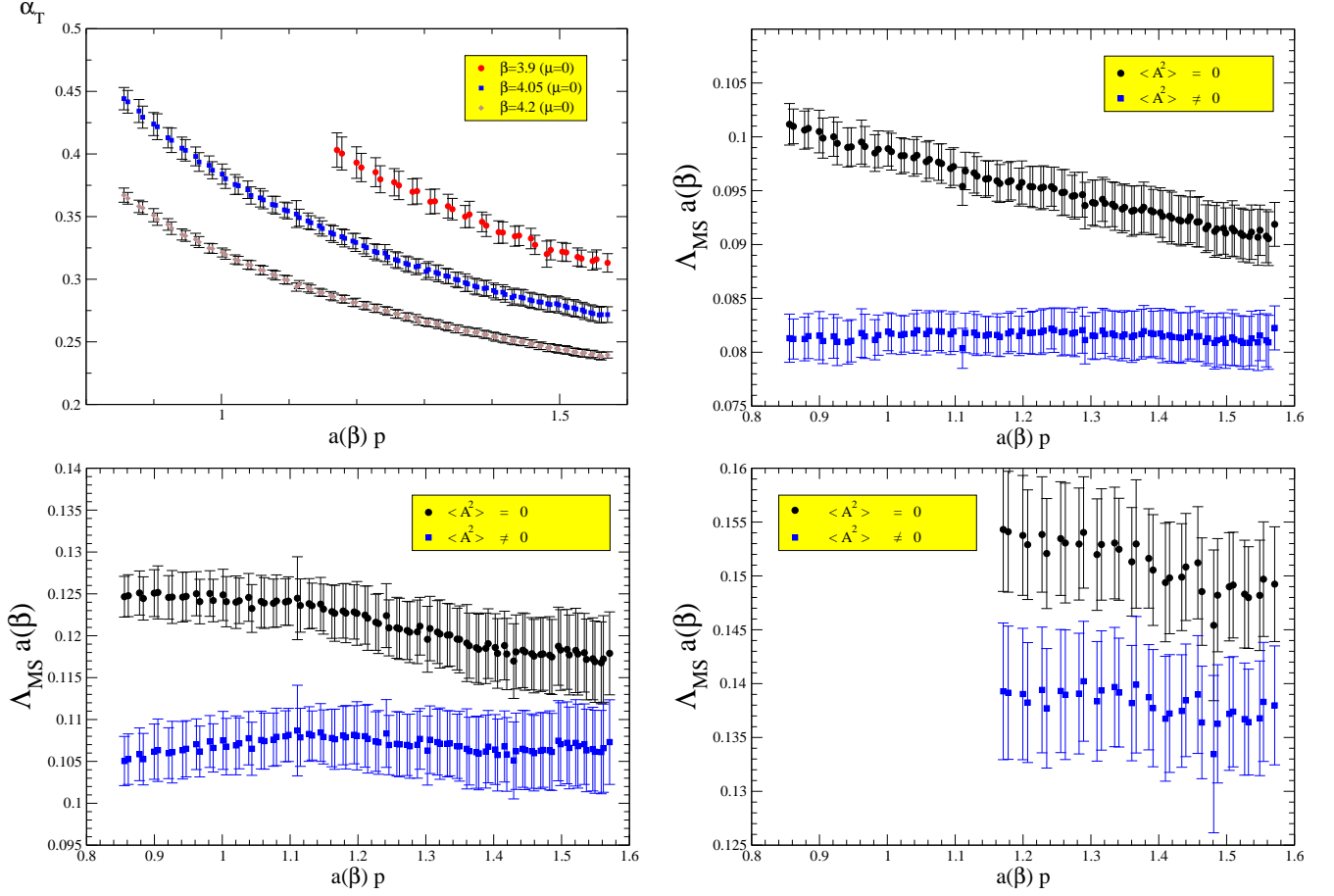

\begin{center}
\begin{tabular}{cc}
\includegraphics[width=8.2cm]{figs/alpha.eps} 
&
\includegraphics[width=8.5cm]{figs/Lambdab42.eps} 
\\
\includegraphics[width=8.5cm]{figs/Lambdab405.eps}
&
\includegraphics[width=8.5cm]{figs/Lambdab39.eps}
\end{tabular}
\end{center}
\caption{\small Left-hand plot (up): The Taylor coupling, free of $H(4)$ and
mass-quarks artefacts, computed by  applying \eq{alpha:mu} for the three
$\beta=3.9,4.05,4.2$ and plotted in terms of  the lattice momentum $a(\beta) p$.
The three other plots show $\Lambda_{\overline{\rm MS}}$ in lattice units, 
computed by the inversion of \eq{eq:invert} with the lattice couplings plotted
in  the upper left-hand plot, for $\beta=4.2$ (upper right), $\beta=4.05$
(bottom left)  and $\beta=3.9$ (bottom right); the black circles are for a
perturbative inversion (with zero gluon condensate) and blue squares are
computed with the best-fit of the gluon condensate (see the text).}
\label{fig:alpha}
\end{figure}

The analysis of the slopes (see fig.~\ref{fig:C0}) leaves us with both a fair
estimate of $R_0$ above 2.8 GeV, $R_0 \sim -0.9 \cdot 10^2$, and an intrinsic definition for the momentum window
where \eq{alpha:mu} can be applied  to extrapolate at any momentum. Thus, after
the extrapolation with the previously obtained $R_0$ down to vanishing $a\mu_q$, one
obtains the three estimates for the running coupling at $\beta=3.9,4.05,4.2$
plotted  in terms of the momentum in lattice units, $a p$, in
fig.~\ref{fig:alpha}.(a), which shows a very smooth running behaviour. These are
the  lattice estimates for the coupling to be confronted to the analytical 
prediction given by \eq{alphahNP}. 

 The fact that $R_0$ with our present data  goes to the same constant  for both
$\beta$'s, leads us to consider that the $\mu_q$ dependence of $\alpha$ is mainly 
  a lattice artefact (else it should be a
function of $\mu_q$ and not of $a\mu_q$). The slope  $R_0$ is not small. We did not
expect this. It has to be seriously considered as it affects the result on
$\Lams$. The slope being negative, the extrapolation to vanishing  $a\mu_q$
leads to a larger value for $\Lams$ than if we had estimated it at finite
$a\mu_q$. We shall see that this effect is of the order of an increase of 40 MeV
on $\Lams$.

\vspace*{0.1cm}

{\it Taking into account the effects due to dynamical quarks in a global
analysis of the  lattice determinations is among the main results of this paper.
This  will lead to a proper extrapolation to the continuum limit.}

\section{Computing $\Lambda_{\overline{\rm MS}}$ and the gluon condensate}
\label{sec:anal}

In this section, following~\cite{Boucaud:2008gn}, we will apply a ``{\it
plateau}''-procedure  exploiting \eq{alphahNP} to get a reliable estimate of
the  $\Lambda_{\rm QCD}$-parameter from the Taylor running coupling constant
computed as explained in  the previous section from lattice simulations with
$N_f=2$ dynamical quark flavors. 

\subsection{Looking for the ``{\it plateau}''}

In fig.~\ref{fig:alpha}, we show the estimates of $\Lambda_{\overline{\rm MS}}$ 
obtained when interpreting the lattice coupling computed from \eq{alphaT} for 
$\beta=3.9,4.05,4.2$ for all momenta inside the window where the slope for the 
$\mu_q$-extrapolation behaves as a constant, i.e.  for $ p \ge p_{\rm min} 
\simeq 2.8$ GeV, up to our chosen lattice upper bound~\footnote{Above some value of $ap$
the lattice artefacts become overwhelming. One must choose an upper bound, this
choice being to some extent arbitrary.}  of $a p < 1.6 $. The estimate of
$\Lams$ is done first (black circles) thanks to the {\it inverted}   four-loop perturbative formula for
the coupling, \eq{betainvert}. These estimates systematically  decrease as the
lattice momentum increases, while if we were in the perturbative
region it should be a constant as $\Lams$ is, by definition, a constant in the
perturbative expansion.  This clearly reveals the necessity of applying the
non-perturbative formula including power  corrections, \eq{alphahNP}, with a
non-zero gluon condensate.  This is also done in fig.~\ref{fig:alpha}, where the
same is plotted but inverting  instead the non-perturbative formula (blue
squares).  The value of the gluon condensate has been determined by requiring  a
``{\it plateau}'' to exist over the total momenta window.  More precisely, one
requires the best-fit to a constant of the estimates of  $\Lambda_{\overline{\rm
MS}}$, in lattice units, in terms of the lattice momentum,
\beq
(x_i,y_i) &\equiv& \left( a p_i,\Lambda_i \right) \ , 
\eeq
where, of course, $i$ runs to cover all the lattice estimates of the 
coupling inside the defined window and where $\Lambda_i$ is obtained by 
inverting~\footnote{For the sake of simplification we are using here  the tree level 
value for the Wilson coefficient of the dimension-2 condensate, i.e. a constant ;  
higher orders will be considered in the next sections.} 
\beq\label{eq:invert}
\alpha_T^{\rm pert}\left(\log\frac{a^2 p^2_i}{\Lambda_i^2}\right) &=& 
\frac{\widehat{\alpha}_T(a^2 p^2_i,0)}{\displaystyle 1+\frac{c}{a^2 p^2_i}} \ ,
\eeq
with $\alpha_T^{\rm pert}$ given by the perturbative four-loop formula
\eq{betainvert},  $\widehat{\alpha}_T(a^2 p^2_i,0)$ taken from the extrapolation
of  the lattice couplings by \eq{alpha:mu} and $c$ resulting from the best-fit 
(it appeared written in terms of the gluon condensate in \eq{alphahNPt}~) of
$\Lambda_i$  to a constant. Thus, $\Lambda_i$ is required to reach a ``{\it
plateau}'', behaving  as a constant when $i$ runs, in terms of the lattice
momentum.

This procedure leaves us with estimates of $\Lambda_{\overline{\rm MS}}$ and the 
gluon condensate (computed from the best-fit determination of $c$), expressed 
in lattice units, for any $\beta$. However, we will take this as a striking 
illustration of the necessity of including non-perturbative power corrections 
in the analysis (see fig.~\ref{fig:alpha}) but we will only report the 
results obtained in the next section when a global fitting strategy will be 
applied to our $\mu_q$-extrapolated lattice data for 
the three different $\beta$'s.

\subsection{Global fit and the calibration of lattice spacing}

The running of $\alpha_T$ given by the combination of Green
functions in eq.~(\ref{alphaT}) and the extrapolation through \eq{alpha:mu}, provided that 
we are not far from the continuum limit and discretization errors are treated properly, 
depend only on the momentum (except, maybe, finite volume errors at low momenta).
The supposed scaling of the Taylor coupling implies for the three curves plotted in 
fig.~\ref{fig:alpha}.(a) to match to each other after
the appropriate conversion of the momentum (in x-axis) from lattice to 
physical units, with the multiplication by the lattice spacing at each $\beta$. 
Thus, we can apply the ``plateau''-method described in the previous subsection 
for the three $\beta$'s all at once by requiring the minimisation of the total $\chi^2$:
\beq\label{chi2T}
\chi^2\left(a(\beta_0)\Lams,c,\frac{a(\beta_1)}{a(\beta_0)},\frac{a(\beta_2)}{a(\beta_0)}\right) \ = \ \sum_{j=0}^2 \sum_{i} \
\frac{\left( \Lambda_i(\beta_j) - \displaystyle \frac{a(\beta_j)}{a(\beta_0)} a(\beta_0) \Lams  \right)^2}
{\delta^2(\Lambda_i)} \ ;
\eeq
where the sum over $j$ covers the sets of coupling estimates for the three $\beta$'s 
($\beta_0=3.9$, $\beta_1=4.05$, $\beta_2=4.2$), the index $i$ 
runs to cover the fitting window defined, as previously explained, through the 
slope analysis~\footnote{In the case of $\beta_2=4.2$, as only the simulation for one  
quark mass, $\mu_q=0.0065$, is exploited, one extrapolates by applying the slope $R_0$ 
computed for $\beta_0=3.9$ and $\beta_1=4.05$ all the coupling estimates obtained 
inside the same lattice-momentum window determined for $\beta_1=4.05$. So we do 
because simulations at both $\beta_1=4.05$ and $\beta_2=4.2$ where performed in 
$32^3\times64$ lattices and the impact of volume effects were supposed to determine 
the lower bound of the fitting window.} and $\Lambda_i(\beta_j)$ is again obtained 
for any $\beta_j$ by requiring
\beq
\alpha_T^{\rm pert}\left(\log\frac{a^2(\beta_j) p^2_i}{\Lambda_i^2(\beta_j)}\right) 
\ 
\left( 1+\frac{c}{a^2(\beta_j) p^2_i} 
\left(\frac{a(\beta_j)}{a(\beta_0)}\right)^2 
\left(\frac{\log\frac{a^2(\beta_j) p^2_i}{\Lambda_i^2(\beta_j)}}
{\log\frac{a^2(\beta_0) q_0^2}{\Lambda_i^2(\beta_j)}} \right)^{-\frac{27}{116}}
\right)
\ = \ 
\widehat{\alpha}_T(a^2(\beta_j) p^2_i,0) 
\eeq
where now we apply the OPE formula including the leading logarithm correction for 
the Wilson coefficient, \eq{alphahNP}, 
with $\alpha_T^{\rm pert}$ given by the perturbative four-loop formula, \eq{betainvert}, 
and where $ a(\beta_0) q_0=4.5$ (this means $q_0 \approx 10$ GeV). 
The errors for the extrapolated couplings are estimated by jackknife analysis and properly 
propagated through the perturbative inversion to give $\delta(\Lambda_i)$.
The function $\chi^2$ is minimised over the functional space defined by the 
four parameters that are explicitly put in arguments for \eq{chi2T}'s l.h.s.: 
$a(\beta_0)\Lams$, $c$, $\frac{a(\beta_1)}{a(\beta_0)}$, $\frac{a(\beta_2)}{a(\beta_0)}$.
Thus we obtain all at once $\Lams$ and the gluon condensate, in units of the 
lattice spacing for $\beta_0=3.9$, and the ratios of lattice spacings for our three
simulations after the extrapolation to the limit $\mu_q \to 0$ (see tab.~\ref{tab:resF}).
The errors are calculated again by jackknife analysis.

\begin{figure}[hbt!]
\begin{center}
\includegraphics[width=15cm,height=10cm]{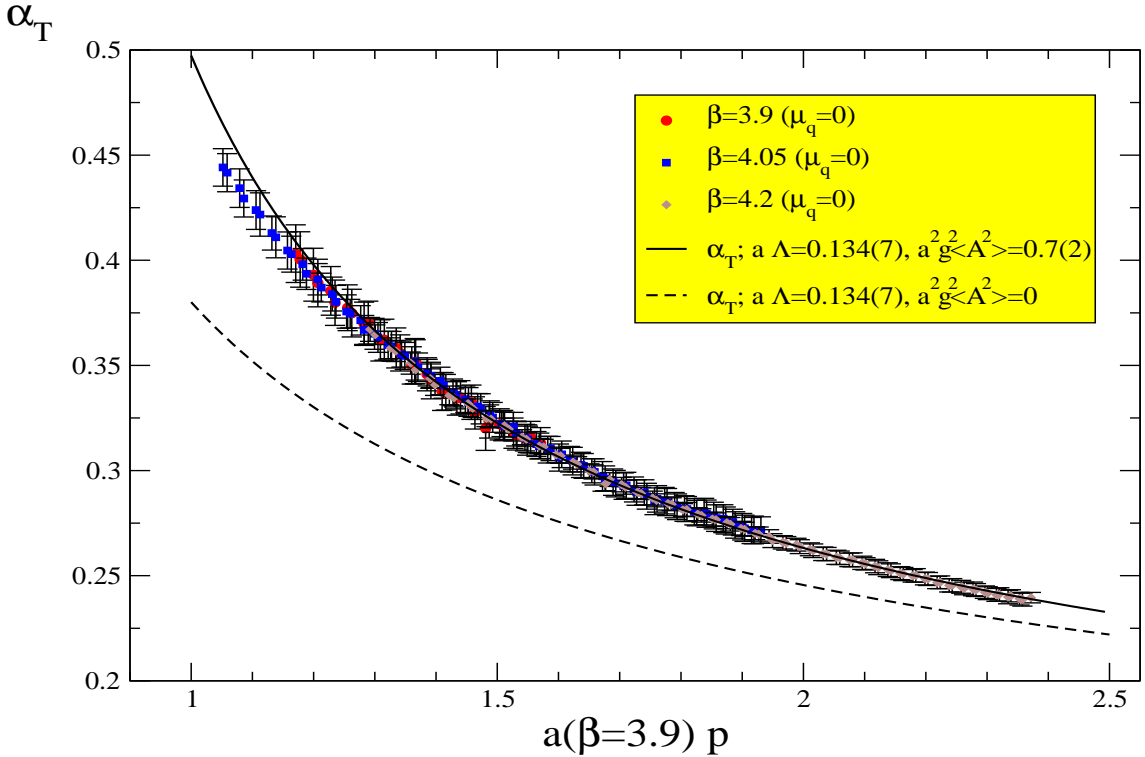}
\end{center}
\caption{\small The scaling of the Taylor coupling computed by 
applying \eq{alpha:mu} for the three $\beta=3.9,4.05,4.2$ is shown. The lattice momentum, 
$a(\beta)p$ in the x-axis, is converted to a physical momentum in units 
(the same for the three $\beta$'s) of $a(3.9)^{-1}$. Fo so to do, the ratios of the lattice spacings  
must be applied. The solid curve is the non-perturbative prediction given by
\eq{alphahNP} with the best-fit parameters for $\Lams$  and the gluon condensate, 
and the dotted one is the same but with zero gluon condensate.}
\label{fig:alphaR}
\end{figure}

The ratios of lattice spacings can be applied to express the momenta 
for all the three sets of coupling estimates plotted in fig.~\ref{fig:alpha} (upper left-handed plot) 
in units of the lattice spacing at $\beta=3.9$. Thus they indeed match each other 
and fit pretty well to the analytical prediction given by \eq{alphahNP} with the 
best-fit parameters for $\Lams$ and the gluon condensate, in units of $1/a(3.9)$ 
(see tab.~\ref{tab:resF}), as can be seen in the plot 
of fig.~\ref{fig:alphaR}~. The quality for the fits drastically deteriorates as data below
$a(3.9) p \simeq 1.2$ are included, whichever value results for the gluon condensate. 
Thus, a fitting window excluding those data are applied in obtaining the best-fit parameters 
in tab.~\ref{tab:resF} and the best-fit curve in fig.~\ref{fig:alphaR} which 
clearly detaches from the data below such a lower bound.  

\begin{table}[hbt!]
\begin{center}
\begin{tabular}{|c|c|c|}
\hline
 & This paper & String tension \\
\hline
$a(3.9)/a(4.05)$ & 1.224(23) & 1.255(42)  \\
\hline
$a(3.9)/a(4.2)$ & 1.510(32)& 1.558(52)  \\
\hline
$a(4.05)/a(4.2)$ & 1.233(25)& 1.241(39) \\
\hline
$\Lams a(3.9)$ & 0.134(7) &  \\
\hline
$g^2 \langle A^2 \rangle a^2(3.9)$ & 0.70(23) & \\
\hline 
\end{tabular}
\end{center}
 \caption{\small Best-fit parameters for the ratios of lattice spacings, $\Lams$ 
and the gluon condensate (for which $a(3.9) q_0=4.5$ is chosen). For the sake 
of comparison, we also quote the results from \cite{Baron:2009wt} that 
were obtained by computing the hadronic quantity, $r_0/a(\beta)$, and applying 
to it a chiral extrapolation.}
\label{tab:resF}
\end{table}


\subsection{The contribution from the Wilson coefficient higher orders}

The Wilson coefficients for gluon and ghost propagators have been very recently obtained at four-loop 
level~\cite{Chetyrkin:2009kh}. As it is shown in appendix, by exploiting the results 
of this ref.~\cite{Chetyrkin:2009kh}, the four-loop OPE formula for the T-scheme coupling 
for $N_f=2$ can be obtained by replacing in \eq{alphahNP}:

\beq\label{prescr}
g^2_T(q_0^2) \ &\to& \  
g^2_T(q_0^2) \left(1 + 1.2932 \ \alpha(\mu^2)+1.9363 \ \alpha^2(\mu^2)+ 3.8296 \ \alpha^3(\mu^2) \right)
\nonumber \\
&\times& \left(1 -0.7033 \ \alpha(q_0^2) - 0.3652 \ \alpha^2(q_0^2) + 0.0051 \ \alpha^3(q_0^2) \right)
\eeq
where $\alpha=\alpha^{\rm pert}_T$ and $q_0^2$ is the renormalization momentum for the local 
operator $A^2$ (see next \eq{alphahNPt3} and compare with \eq{alphahNP} that only incorporates 
the leading logarithmic contribution) which we fixed, as previously indicated, by requiring: 
$a(3.9) q_0=4.5$~. 

Then, we can repeat the analysis of previous sections, after the replacement prescribed in 
\eq{prescr}, and obtain the results collected in tab.~\ref{tab:W1234l}. Thus, a strong stability 
results from the fits for the estimates of $\Lams$ (all of them being compatible within the 
statistical uncertainties and varying less than a 2.3 \%) and a fairly convergent behaviour for 
that of the gluon condensate which, computed with the one-loop Wilson coefficient, clearly 
borrows something from next-to-leading contributions.

\begin{table}[hbt!]
\begin{center}
\begin{tabular}{|c|c|c|c|c|}
\hline
& One loop & Two loops & Three loops & Four loops \\
\hline
$\Lams a(3.9)$ & 0.134(7) & 0.136(7) & 0.137(7) & 0.138(7) \\
\hline
$g^2 \langle A^2 \rangle a^2(3.9)$ & 0.70(23) &  0.52(18) & 0.44(14) & 0.39(14) \\
\hline 
\end{tabular}
\end{center}

 \caption{\small Best-fit parameters for $\Lams$ 
and the gluon condensate (for which $a(3.9) q_0=4.5$ is chosen) by applying a OPE formula including 
the logarithmic corrections for the Wilson coefficient at one, two, three and four loop order. No 
difference is seen for the ratios of lattice spacings.}
\label{tab:W1234l}
\end{table}

\subsection{Discussing the systematical uncertainties}

The main sources of systematical uncertainties affecting the determination of
the  best-fit parameters ($\Lams$, the non-perturbative gluon condensate  and
the ratios of lattice spacings) of the matching previously described are
expected to  come from the truncation of the perturbative series for the
theoretical prediction of  the coupling in \eq{alphahNP}, the possible effect of
higher-orders  in the OPE expansion and from the finite volume effects in
lattice simulations.  We will pay attention in the following to these error
sources.

\subsubsection{Volume effects}

As can be seen in tab.~\ref{setup}, we exploited lattice simulations in volumes 
$24^3\times 48$ at  $\beta=3.9$ and 
$32^3\times 64$ at $\beta=4.05$ and $4.2$.  
In the case of $\beta=4.2$, in order to spare computing time, we use a volume smaller than what
is usually needed to measure hadronic quantities (in particular, a $48^3\times 96$ 
lattice at $\beta=4.2$ and $a\mu_q=0.002$ is required), relying on the hope
that, being interested in ultraviolet quantities, the finite volume effects will
be reduced. One expects that the smaller the product of momentum and lattice
size, the larger is  the volume-effect impact. Indeed, we introduced a lattice
momentum cut, $a(\beta) p_{\rm min}$,  when we studied the quark-mass
extrapolation, which the slopes for the mass  squared extrapolation detached
below from the constant behaviour, and we interpreted this  as a possible volume
effect. Furthermore, in fig.~\ref{fig:alphaR}, no impact of any remaining 
volume effect on the determination of the coupling is seen: the impressive
scaling shown by the results from our three simulations  in fig~\ref{fig:alphaR}
seems to confirm that we are finally  left with no important volume effect.

\subsubsection{Three-loop versus four-loop confrontation}

A standard way to estimate the effect of perturbative-series truncation is 
to repeat the analysis described in previous sections but applying instead 
a three-loop formula for the perturbative inversion. 
If this is done, one obtains the results collected in tab.~\ref{tab:3l4l}. 


\begin{table}[hbt!]
\begin{center}
\begin{tabular}{|c|c|c|}
\hline
& Four loops & Three loops \\
\hline
$a(3.9)/a(4.05)$ & 1.224(23) & 1.229(23) \\
\hline
$a(3.9)/a(4.2)$ & 1.510(32)& 1.510(29)  \\
\hline
$a(4.05)/a(4.2)$ & 1.233(26)& 1.234(25)  \\
\hline
$\Lams a(3.9)$ & 0.134(7) & 0.125(6) \\
\hline
$g^2 \langle A^2 \rangle a^2(3.9)$ & 0.70(23) & 0.80(20) \\
\hline 
\end{tabular}
\end{center}
 \caption{\small Best-fit parameters for the ratios of lattice spacings, $\Lams$ 
and the gluon condensate (for which $a(3.9) q_0=4.5$ is chosen).}
\label{tab:3l4l}
\end{table}

Then, we can conclude that no noticeable impact from the perturbative truncation 
is resulting on the determination of the ratios of lattice spacing. Nevertheless, 
a systematic uncertainty of roughly a 7 \% can be estimated from the discrepancy of 
the three and four-loops estimates for the $\Lambda_{\overline{\rm MS}}$. 
Analogously, the determination of the gluon 
condensate is affected by a correction of the order of 13 \%. 

\subsubsection{The impact of higher-orders in the OPE expansion}

We previously paid attention to the comparison of the best-fit results 
when applying Wilson coefficients for the OPE expansion at different loop-orders
(see tab.~\ref{tab:W1234l}). We then concluded that the different estimates of $\Lams$ 
differ from each other less than a 2.3 \%, while those for the gluon condensate 
fairly converge as the loop-order increases to roughly one half of the 
one-loop result.

On the other hand, as previously explained, the quality for the fits 
drastically deteriorates as data below $a(3.9)p \simeq 1.2$ are considered  (the
$\chi^2$ becomes of the order of four times larger).  This could be an
indication of the impact of OPE higher power corrections that could be  simply
parameterized as
\beq\label{ope4}
\alpha_{T,P4}(\mu^2) \ = \ \alpha_T(\mu^2) + \frac{c_4}{\mu^4} \ ,
\eeq
where $\alpha_T$ is given by \eq{alphahNP} and $c_4$ is a constant encoding all
the information coming  from the condensates of higher-dimension operators  and
the Wilson coefficients (their anomalous dimension is thus neglected).  If we
try a fit with \eq{ope4} to the lattice data,  good-quality fits are obtained 
for negative values of $c_4$, of the order of $-0.1$, while the positive
estimated contribution of the gluon condensate increases drastically. The
fitting function with or without the ${c_4}/{\mu^4}$ turn out to be very close
to one another over the whole momentum window. In other words there is a valley
in the parameter space in which both the $1/\mu^4$ and the $1/\mu^2$
coefficients vary in an anticorrelated way without any significant change of the value of 
fitting function. Furthermore, including the ${c_4}/{\mu^4}$ term increases
drastically the errors. $\Lams$ is also correlated with the other
two parameters: it decreases when ${c_4}/{\mu^4}$ decreases below zero, by about 10 \% from the fit without  ${c_4}/{\mu^4}$ term and the fit with it.

Our conclusion is that the fit with ${c_4}/{\mu^4}$ is extremely unstable.
We therefore decide not to include it in our fits. The resulting systematic
uncertainties is not larger than $\sim 10 \%$ on $\Lams$. The estimate of the
gluon condensate may be more severely affected. But, as shown in the quenched
case~\cite{Boucaud:2008gn}, the estimates of the gluon condensate stemming from
different quantities, when neglecting ${c_4}/{\mu^4}$ terms, are quite
compatible. This would not be possible if the dimension-four operators, with
different Wilson coefficients for every quantity, were playing a significant 
role. We remain thus rather confident in our estimate of $<A^2>$.
 What should be stressed here is that {\it the necessity of a
positive non perturbative contribution is unavoidable, and that it is well taken
into account by the dominant dimension-two $<A^2>$ condensate.}


\subsection{Conversion to physical units and quark mass effect}

In this section, we will apply the results for the lattice spacings at $\beta=3.9$, $4.05$ and $4.2$ obtained 
in ref.~\cite{Baron:2009wt} through a very exhaustive investigation of the light meson physics using 
maximally twisted mass fermions for $N_f=2$ degenerated quark flavours (the gauge configurations we use in 
this work were part of the data ensembles generated by ETMC and analysed in ref.~\cite{Baron:2009wt}).  
The physical scale is given by requiring $f_{\pi}=130.7$ MeV as also done in \cite{Boucaud:2007uk}. 
We first computed the ratios of those lattice spacings obtained in ref.~\cite{Baron:2009wt} 
and compared them with the ones obtained in this paper in tab.~\ref{tab:resF}. The agreement is indeed 
remarkable. Then, by applying the result~\cite{Baron:2009wt}
\beq\label{cal}
a(3.9) \ = \ 0.0801(14)~\mbox{\rm fm} \ ,
\eeq
to convert into physical units our estimates of $\Lams$ and the gluon condensate (see tab.~\ref{tab:resF}), 
the final results are:
\beq\label{final}
\Lams &=& \left(330 \pm 23 \pm 22 _{-33}\right) 
\times \left(\frac{0.0801 \ \mbox{\rm fm} \cdot  130.7 \ \mbox{\rm MeV} \ }{a(3.9) \ f_\pi} 
\right)  \mbox{\rm MeV} \ , \nonumber \\ 
g^2(q_0^2) \langle A^2 \rangle_{q_0} &=&  \left(4.2 \pm 1.5 \pm 0.7 ^{+ ?}\right) \times 
\left( \frac{0.0801 \ \mbox{\rm fm} \cdot  130.7 \ \mbox{\rm MeV} \ }{a(3.9) \ f_\pi} 
\right)^2 \
\ \mbox{\rm GeV}^2 \ ;
\eeq
where we first quote the purely statistical errors, where the one from the lattice size 
determination in ref.~\cite{Baron:2009wt} has been properly into account, and then 
the main systematical uncertainties which were detailed in the previous subsection and 
that, as explained, mainly tend to increase the gluon condensate and to decrease the value of 
$\Lams$. 
In particular, for the gluon condensate, the contribution to the uncertainty of higher 
powers in OPE expansion, although unequivocally increasing its size, is very hard to 
be estimated (as discussed in the previous section) and we explicitely indicated this 
by the addition of a question mark in the upper systematical errors of \eq{final}. 
It should be also remembered that $q_0= 4.5 \ a(3.9)^{-1} = 11.1$ GeV. 
In ref.~\cite{Baron:2009wt}, the lattice spacings 
result from combined fits including several lattice simulations at 
different $\beta$'s, after chiral extrapolations on the quark mass, where the physical 
scale is fixed by $f_\pi = 130.7$ MeV. In particular, \eq{cal} and the ratios 
we presented for the sake of comparison 
in tab.~\ref{tab:resF}, were obtained in that ref.~\cite{Baron:2009wt} through 
a combined fit including $\beta=3.9,4.05$ and $4.2$, but results obtained through 
combined fits including either $\beta=3.8,3.9$ or $\beta=3.8,3.9$ and $4.05$
were also reported. Had we applied instead of \eq{cal} the other results reported 
in~\cite{Baron:2009wt} in order to convert our best-fit parameters into physical 
units, we would obtain the results of tab.~\ref{tab:calib}. Thus, we can roughly 
estimate how our final result in \eq{final} is systematically affected 
by the conversion to physical units and consider the central value of $\Lams$ 
roughly to range from $313$ to $335$ MeV and that of the gluon condensate 
from $3.8$ to $4.3$ GeV$^2$. 
Furthermore, in order to make easier any further comparison (avoiding also the ambiguities 
related to the conversion to physical units), we collect in tab.~\ref{tab:less} the ratios 
of $\Lams$ and some momentum-dimension physical quantities also obtained 
in ref.~\cite{Baron:2009wt}. In particular, if we compare our $N_f=2$ estimate 
for $\Lams$, converted to physical units, in \eq{final} and the same obtained by applying 
the Sch\"odinger functional method in ref.~\cite{DellaMorte:2004bc}, $\Lams=245(16)(16)$ MeV, 
they clearly differ. Nevertheless, had we compared their estimate of $r_0 \Lams=0.62(4)$ with 
ours in tab.~\ref{tab:less}, $r_0 \Lams=0.72(5)$, we would conclude that they almost agree 
with each other within their statistical error intervals. Thus, this indicates that the main 
source of discrepancy for these two results comes from setting the physical scale.

\begin{table}[hbt!]
\begin{center}
\begin{tabular}{|c|c|c|c|}
\hline
$\beta$'s in fits & $a(3.9)$ (fm) & $\Lams$ (MeV)& $g^2 \langle A^2 \rangle$ (GeV$^2$) \\
\hline
3.9, 4.05, 4.2 & 0.0801(14) & 330(23) & 4.2(1.5) \\
\hline
3.9, 4.05 & 0.0790(26) & 335(28) & 4.3(1.7) \\
3.8, 3.9, 4.05 & 0.0847(15) & 313(22) & 3.8(1.4)   \\
\hline 
\end{tabular}
\end{center}
 \caption{\small Best-fit parameters for $\Lams$ 
and the gluon condensate (for which $a(3.9) q_0=4.5$ is chosen) by applying the different 
lattice spacings for $a(3.9)$ obtained in ref.~\cite{Baron:2009wt} after combined fits 
including simulations with the different $\beta$'s indicated in the first column.}
\label{tab:calib}
\end{table}

\begin{table}[hbt!]
\begin{center}
\begin{tabular}{|c|c|c|c|c|}
\hline
 & $f_\pi$ & $f_0$ &$1/r_0$ & $m_{u,d}$ \\
\hline 
$\Lams$ & 2.52(18) & 2.71(19) & 0.72(5) & 92(9) \\
\hline 
\end{tabular}
\end{center}
 \caption{\small dimensionless ratios of $\Lams$ and some momentum dimension quantities taken 
from ref.~\cite{Baron:2009wt}, where the gauge configurations 
analyzed in this paper were also exploited. Each number in the table is obtained 
by dividing the quantity indicated in the row label ($\Lams$) over the one indicated in the 
column label.}
\label{tab:less}
\end{table}

\begin{figure}[hbt]
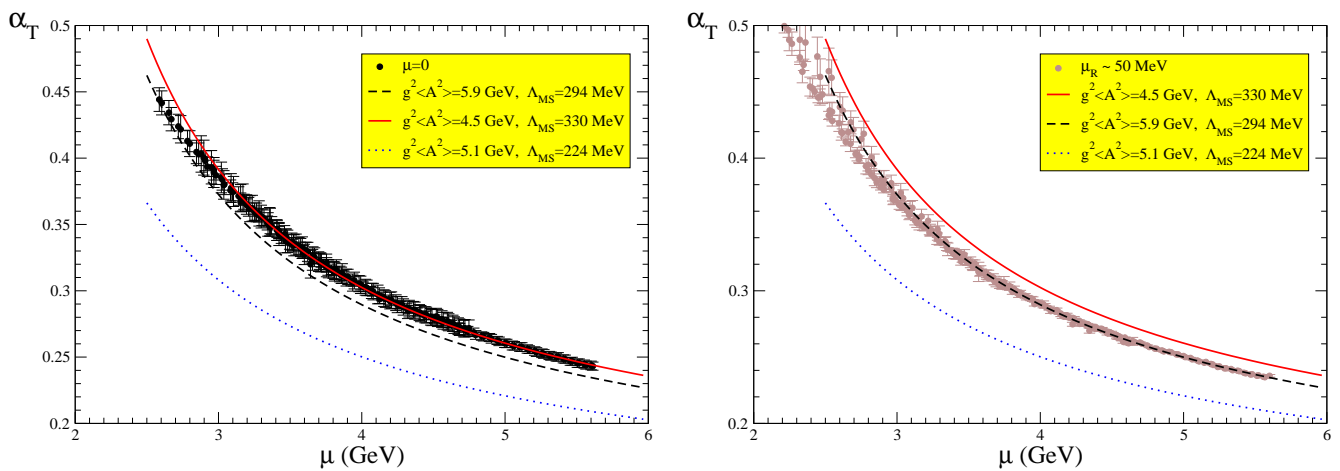

\vspace*{0.42cm}
\begin{center}
\begin{tabular}{cc}
\includegraphics[width=8.5cm]{figs/alphaGEV.eps}
&
\includegraphics[width=8.5cm]{figs/alphaGEVmu50.eps}
\end{tabular}
\end{center}
\caption{\small The Taylor coupling after the chiral extrapolation plotted 
in fig.~\ref{fig:alphaR}, but now in terms of the momentum in physical units (left),  
and the same Taylor coupling computed for the three lattice data sets of tab.~\ref{tab:masses}, 
without chiral extrapolation, also in terms of the momentum in physical units (right).
The OPE formula with the best-fit parameters obtained for extrapolated lattice 
data at $\mu_R=0$ (solid red line), for the lattice data at $\mu_R \sim 50$ MeV (black dashed line)  
and for the quenched ($\mu_q \to \infty$) lattice data in ref.~\cite{Boucaud:2008gn} 
(blue dotted line) are displayed in both plots for the sake of comparison. 
}
\label{fig:comp}
\end{figure}

On the other hand, for the purpose of illustrating the effects derived from the dynamical 
quark mass on the determination of $\Lams$, we can analyze instead of the $\mu_q$-extrapolated 
data for the coupling (plotted in fig.~\ref{fig:alpha}) the ones obtained from 
the the lattices at $\beta=4.2$ with $a\mu_q=0.0065$, $\beta=4.05$ with $a \mu_q=0.008$ and 
$\beta=3.9$ with $a\mu_q=0.010$. In view of the results of the lattice spacings ratios in 
tab.~\ref{tab:resF} and \eq{cal}, these assumed to be independent of the quark mass, and after 
the appropriate renormalization, 
\beq
\mu_R \ = \ \frac{\mu_q}{Z_P(q_0)} \ ,
\eeq
where we apply the $\overline{\rm MS}$ renormalization constant, $Z_P$, at the renormalization scale $q_0=2$ GeV 
given in ref.~\cite{Baron:2009wt}, three very similar quark masses will be obtained for the three 
simulations (see  tab.~\ref{tab:masses}). Then, by implementing the 
``plateau method'' with the OPE formula in \eq{alphahNP}, but applying \eq{cal} 
to convert into physical units, one will be left with an estimate of $\Lams$ at a 
renormalized quark mass of the order of $\mu_R \simeq 50$ MeV (see fig.~\ref{fig:comp}). 
The result of this analysis is:
\beq\label{eq:mu50}
\Lams \ = \ 294 \pm 10 \mbox{\rm MeV} \ ,
\eeq  
that appears to indicate a trend that should be kept in mind to compare with 
previous (or future) unqueched estimates of $\Lams$: the larger is the quark
mass,  the lower is the estimate of $\Lams$. For instance, in
ref.~\cite{Boucaud:2001qz}, $\Lams=264$ MeV  was reported as the result of a
preliminary analysis of the three-gluon coupling from lattice simulations with 
two flavours of Wilson dynamical quarks and with a renormalized sea-quark mass
roughly ranging  from 100 to 400 MeV. Both this preliminary results and that of
\eq{eq:mu50} seem to be in the right ballpark.   So more if one considers the
uncertainty derived from the physical lattice calibration previously discussed 
(see tab.~\ref{tab:calib}) and which makes the central value of $\Lams$ to range
from 278 to 298 MeV.

\begin{table}[hbt!]
\begin{center}
\begin{tabular}{|c|c|c|}
\hline
 $\beta$ & $a \mu_q$ & $\mu_R$ (MeV)  \\
\hline
4.2 & 0.0065 & 53.4 \\
\hline
4.05 & 0.008 & 54.9 \\
\hline
3.9 & 0.010 & 52.8 \\
\hline
\end{tabular}
\end{center}
 \caption{\small Quark masses for the three simulations which we re-analyze the results from 
in order to get an indication of the quark mass effect.}
\label{tab:masses}
\end{table}

This trend for the behaviour of the estimate for $\Lams$ from the lattice, in particular from 
the lattice strong coupling in Taylor scheme, is corroborated by the results obtained from 
the analysis of quenched lattice simulations in ref.~\cite{Boucaud:2008gn} 
(where $\Lams=224$ MeV). This can be clearly seen in the plots of fig.~\ref{fig:comp}, where 
the OPE formulae with the values of $\Lams$ and the gluon condensate obtained  
by fits of results from lattices at $\mu_R=0$, $\mu_R \sim 50$ in this paper and by a fit 
of results from quenched lattices in ref.~\cite{Boucaud:2008gn} appear displayed. It should be  
noted that the quenching approximation can be understood as equivalent to consider infinite dynamical 
quark masses. It may be also worth to remember that, in the three cases, the OPE formulae 
fit pretty well to the lattice results for the Taylor coupling from momenta roughly 
ranging from 3 to 6 GeV.

\section{Discussions and conclusions}
\label{sec:conclu}

We analyzed gauge configurations with twisted $N_f=2$ dynamical quarks at the
{\it maximal twist}  produced by the ETM collaboration for lattices at several
$\beta$'s (3.9, 4.05, 4.2), and with several different  bare twisted masses, and
computed the strong coupling constant renormalized in the  so-called Taylor
scheme. The main advantage of this scheme being not to require any three-point
Green function  computation, statistical fluctuations became under enough
control to permit a very elaborate treatment of the lattice artefacts and a
precise estimate of the couplings at the infinite cut-off limit. The coupling
estimates for lattices at different $\beta$'s were seen to match pretty well, as
should happen  if the cut-off limit is properly taken, when plotted in terms of
the renormalization momenta converted  to the same units by applying the
appropriate lattice spacings ratios. These ratios could be either taken  from
independent computations or obtained by requiring the best matching with pretty
compatible results. Indeed, to require the best matching for the strong coupling
computed in terms of the momentum for lattices  at different $\beta$'s was
proposed as an alternative method to determine the lattice spacings ratios  in
ref.~\cite{Boucaud:2008gn} and shown here to work pretty well. 

Thus, once we are left with the estimates of the coupling constant extrapolated
at vanishing dynamical mass $\mu_q$, for every value   of the renormalization
momentum, $\mu$, they were converted via a fit with a four loops formula into
the  value of $\Lams$. As also noticed for the analysis of lattice results with
$N_f=0$, the values of $\Lams$  so obtained for every value of $\mu$ resulted
not to be independent on $\mu$ over a momentum window roughly  ranging from 2 to
6 GeV. This implies a non-negligible impact of non-perturbative contributions
which  we accounted for by a Wilson OPE expansion assumed to be dominated by the
condensate of the dimension two  operator: $A^2$. Thus, after including the
gluon condensate contribution we recovered a ``plateau'' for  the fitted values
of $\Lams$ in terms of the renormalization momentum and estimated  the
condensate.  After discussing the several main sources of systematic
uncertainties, specially the  higher order contributions in both perturbative
and OPE expansions, we converted our fitted parameters into  physical units by
applying\cite{Baron:2009wt}: $a(3.9)=0.0801(14)$ fm. Thus we obtain 330(23)(22)$_{-33}$ MeVs
for $\Lams$ and  4.2(1.5)(0.7) GeV$^2$ for the gluon condensate. As was proven in the
quenched case~\cite{Boucaud:2008gn}, a positive dimension two gluon  condensate
is unequivocally needed to account for the momentum behaviour of the strong
coupling computed from  the lattice. Whether condensates of higher order
operators have to be included or not in the Wilson expansion  is a different
matter that cannot be properly addressed with our current data. Indeed,  a
negative contribution of the order $1/p^4$ can borrow something from the $1/p^2$
gluon condensate,  both OPE formulae with and without such a contribution
remaining almost totally indistinguishable within our fit momentum window.
Then, since the fits including $1/p^4$ contributions become unstable and very
noisy, we do not  take them into account. A possible way-out could be to perform
a cross-check by analyzing the running behaviour of  the renormalization
constant of other operators. In particular for the quenched case,  apart from
gluon and ghost operators, the vector part of the quark propagator is studied in
ref.~\cite{Boucaud:2005rm}  and the fitted gluon condensate is found to be
compatible with the one obtained from the ghost-gluon
vertex~\cite{Boucaud:2008gn}.  A program to study systematically the
non-perturbative OPE contributions to the renormalization constant  of quark
operators is in progress~\cite{OPEquarks}. 
%
%

Furthermore, we also paid attention to the effect of the dynamical quark masses.
As we performed an extrapolation to zero bare twisted mass for the lattice
estimates of the strong coupling constant, the $\Lams$  result corresponds to a
world with $N_f=2$ chiral quarks. Nevertheless, for the sake of comparison, we 
applied the same analysis procedure to three sets of lattice configurations
corresponding to practically the  same renormalized dynamical quark mass
(roughly 50 MeVs), with no mass extrapolation,  and estimated a lighter value of
$\Lams=294(10)$ MeV (the central value can also range from 278 to 298 MeV, depending 
on which value we use for the lattice spacing in physical units). This is
compatible with previous preliminary  results~\footnote{The authors of
ref.~\cite{Sternbeck:2007br} also exploit strong coupling computed  on the
lattice and renormalized in Taylor scheme, although they presented this scheme
as a combination  of MOM prescription for the propagators and the $\overline{\rm
MS}$ for the ghost-gluon vertex.} \cite{Boucaud:2001qz,Sternbeck:2007br}. 
We may recall here that the quenched results\cite{Boucaud:2008gn} for $\Lams$
are still lower, ranging mainly around 230 MeV. Since quenching can be
understood as the situation with infinite dynamical masses, one may infer a 
general trend that $\Lams$ increases when the quark mass decreases, with a
finite value both at infinite and vanishing quark mass. 

As a matter of fact, most of the results for $\Lams$ obtained with Wilson
fermions  (around 260-270 MeV; see for instance~\cite{Gockeler:2005rv} and
references therein)  lie below our zero-mass result, but also below the
phenomenogical value which could be obtained,  after the appropriated
conversion, from the experimental world average, 
$\alpha(M_Z)=0.01184(7)$~\cite{Amsler:2008zzb}.   The standard procedure to
convert $\Lams^{Nf}$ to $\alpha_{\overline{\rm MS}}$ at a given scale, 
typically the mass of Z boson, implies the RGE four-loop evolution of the
coupling and the three-loop  matching at the quark thresholds. Provided that the
$\overline{\rm MS}$ running mass of the charm quark  is 1.5 GeV, the conversion
from $\Lams^{Nf=3}$ to $\Lams^{Nf=4}$ implies to evolve  $\alpha_{\overline{\rm
MS}}$ over an energy range where perturbation theory could fail; but the
conversion  from $\Lams^{Nf=2}$ to $\Lams^{Nf=3}$ is however out of the scope of
the above-described standard  conversion procedure. This is why, at present , we
prefer not to compare with  phenomenological results. This task is to be
properly acomplished when $N_f=2+1+1$ and $N_f=4$ lattice  simulations will be
available. Neverteless, our current results for $\Lams$ with two light quarks
flavours,  after chiral extrapolation, seems to point that systematic effects
due to the dynamical quark masses could  explain the discrepancy between the
lattice estimates for the strong coupling with Wilson fermions  and its
experimental determination.

\section*{Acknowledgements} 
We are particularly indebted to A. Le Yaouanc, J. P. Leroy 
and J. Micheli for participating in many fruitful discussions at 
the preliminar stages of this work.
We also thank the IN2P3 Computing Center 
(Lyon) and the apeNEXT computing laboratory (Rome)
where part of our simulations have been done. 
J. R-Q is indebted to the Spanish MICINN for the 
support by the research project FPA2009-10773 and 
to ``Junta de Andalucia'' by P07FQM02962.

\appendix

\section{Appendix: The Wilson coefficients at the four-loops order}
\label{appendix2}

The purpose of this appendix is to exploit the four-loops results in 
ref.~\cite{Chetyrkin:2009kh} to derive the Wilson coefficients with the 
appropriate renormalization prescription and modify properly \eq{alphahNP}.
Following \cite{OPEone,Boucaud:2008gn} the equations (\ref{OPE1}) for ghost and 
gluon propagators can be rewritten after renormalization as
\beq\label{ap-props}
G_R(q^2,\mu^2) &=& c_0\left(\frac{q^2}{\mu^2},\alpha(\mu^2)\right) + 
c_2\left(\frac{q^2}{\mu^2},\alpha(\mu^2)\right) \ \frac{\langle A^2_R \rangle_\mu}{4 (N_c^2-1) q^2} \ ,
\nonumber  \\
F_R(q^2,\mu^2) &=& \widetilde{c}_0\left(\frac{q^2}{\mu^2},\alpha(\mu^2)\right) + 
\widetilde{c}_2\left(\frac{q^2}{\mu^2},\alpha(\mu^2)\right) \ \frac{\langle A^2_R \rangle_\mu}{4 (N_c^2-1) q^2} \ .
\eeq
With the help of the appropriate renormalization constants, 
one can also write \eq{ap-props} in terms of bare quantities:
\beq\label{ap-bar}
G(q^2,\Lambda^2) &=& Z_3(\mu^2,\Lambda^2) \ c_0\left(\frac{q^2}{\mu^2},\alpha(\mu^2)\right)
\nonumber \\ 
&+&  Z_3(\mu^2,\Lambda^2)  Z_{A^2}^{-1}(\mu^2,\Lambda^2) \ c_2\left(\frac{q^2}{\mu^2},\alpha(\mu^2)\right) \
\frac{\langle A^2 \rangle}{4 (N_c^2-1) q^2} \ ,
\eeq
where $A^2_R=Z^{-1}_{A^2} A^2$.
A totally analogous equation for the ghost dressing function $F(q^2,\Lambda^2)$, 
with $\widetilde{c}_i$ and $\widetilde{Z}_3$ in place of $c_i$ and $Z_3$. 
Now, as the $\mu$-dependence of both l.h.s. and r.h.s. of \eq{ap-bar} should match 
each other for any $q$, one can take the logarithmic derivative with respect to $\mu$ and 
infinite cut-off limit, term by term, on r.h.s. and obtains: 
\beq\label{ap-diffeqs}
\gamma(\alpha(\mu^2)) + 
\left\{ \frac{\partial}{\partial\log\mu^2} 
+ \beta(\alpha(\mu^2))\frac{\partial}{\partial \alpha}\right\} \ 
\ln c_0\left(\frac{q^2}{\mu^2},\alpha(\mu^2)\right) &=& 0 
\nonumber \\
-\gamma_{A^2}(\alpha(\mu^2)) + \gamma(\alpha(\mu^2)) 
+ \left\{ \frac{\partial}{\partial\log\mu^2} 
+ \beta(\alpha(\mu^2))\frac{\partial}{\partial \alpha}\right\} \ 
\ln c_2\left(\frac{q^2}{\mu^2},\alpha(\mu^2)\right) &=& 0 \ ,
\eeq
where $\gamma(\alpha(\mu^2))$ is the gluon propagator anomalous 
dimension and $\gamma_{A^2}(\alpha(\mu^2))$ is the anomalous dimension for the 
local operator $A^2$ defined in \eq{gA2} and that was obtained at four-loop in 
ref.~\cite{Gracey:2002yt}. Both eqs.~(\ref{ap-diffeqs}) can be finally 
combined to give:
\beq\label{ap-fin}
\left\{ -\gamma_{A^2}(\alpha(\mu^2)) +  \frac{\partial}{\partial\log\mu^2} 
+ \beta(\alpha(\mu^2))\frac{\partial}{\partial \alpha}\right\} \ 
\frac{c_2\left(\frac{q^2}{\mu^2},\alpha(\mu^2)\right)}
{c_0\left(\frac{q^2}{\mu^2},\alpha(\mu^2)\right)} \ = \ 0 \ ,
\eeq
and we can proceed in the same way for the ghost dressing function and derive analogous 
equations for the Wilson coefficients, $\widetilde{c}_i$, that differ from those 
for $c_i$ only because $\widetilde{\gamma}(\alpha(\mu^2))$ 
takes the place of $\gamma(\alpha(\mu^2))$. Thus, the combination 
$\widetilde{c}_2/\widetilde{c}_0$ obeys exactly the same \eq{ap-fin}, above derived 
for $c_2/c_0$, that can be solved by applying the following ansatz,
\beq\label{ap-sols}
\frac{\displaystyle c_2\left(\frac{q^2}{\mu^2},\alpha(\mu^2)\right)}
{\displaystyle c_0\left(\frac{q^2}{\mu^2},\alpha(\mu^2)\right)}
\ = \
c_2\left(1,\alpha(q^2)\right) 
\ 
\left( \frac{\alpha(\mu^2)}{\alpha(q^2)}\right)^a \ 
\left( \frac{\displaystyle 1+\sum_i r_i \ \alpha^i(\mu^2)}
{\displaystyle 1+\sum_i r_i \ \alpha^i(q^2)} \right) \ ,
\eeq
where we use that the leading Wilson coefficient is to 
be renormalized in the MOM renormalization prescription such that 
$c_0(1,\alpha(\mu^2))=1$ and where the exponent 
$a$ and the coefficients $r_i$'s are required 
to satisfy \eq{ap-fin}. Concerning the boundary condition for 
$c_2$, the prescription applied for 
the renormalization of the local operator $A^2$ to obtain 
its four-loops anomalous dimension, $\gamma_{A^2}$, in \cite{Gracey:2002yt} is 
the standard $\overline{\rm MS}$ and, with this prescription, 
$c_2$ is computed at the four-loop order in ref.~\cite{Chetyrkin:2009kh} 
(see the eq.~(7) of that paper). We only need to take $q^2=\mu^2$ in 
the expression given in ref.~\cite{Chetyrkin:2009kh} and have thus 
$c_2(1,\alpha(\mu^2))$.
Then, one obtains at the four-loop order:
\beq\label{ap-solGl}
\frac{c_2\left(\frac{q^2}{\mu^2},\alpha(\mu^2)\right)}
{c_0\left(\frac{q^2}{\mu^2},\alpha(\mu^2)\right)}
&=&
3 \overline{g}^2(\mu^2) \left( \frac{\overline{\alpha}(q^2)}
{\overline{\alpha}(\mu^2)}\right)^{\frac{27}{116}} 
\left(1 + 2.1930 \ \overline{\alpha}(q^2) + 6.1460 \ 
\overline{\alpha}^2(q^2)+ 20.5477 \ \overline{\alpha}^3(q^2) \right)
\nonumber \\
&\times& \left(1 + 0.0208 \ \overline{\alpha}(\mu^2) + 0.0095 \ 
\overline{\alpha}^2(\mu^2)+ 0.0164 \ \overline{\alpha}^3(\mu^2) \right)
\ ,
\eeq
where the loop expansion is given in terms of the $\overline{\rm MS}$ coupling, 
$\overline{\alpha}$.

Concerning $\widetilde{c}_2/\widetilde{c}_0$, as it obeys the same differential 
equation \eq{ap-fin}, the solution differ from the one for the gluon propagator 
only because of the boundary condition which will be now obtained from 
eq.~(9) of ref.~\cite{Chetyrkin:2009kh}. Thus, one obtains at the four-loop 
order:
\beq\label{ap-solFh}
\frac{\widetilde{c}_2\left(\frac{q^2}{\mu^2},\alpha(\mu^2)\right)}
{\widetilde{c}_0\left(\frac{q^2}{\mu^2},\alpha(\mu^2)\right)}
&=&
3 \overline{g}^2(\mu^2) \left( \frac{\overline{\alpha}(q^2)}
{\overline{\alpha}(\mu^2)}\right)^{\frac{27}{116}} 
\left(1 +  1.1728 \ \overline{\alpha}(q^2) + 2.7098 \ 
\overline{\alpha}^2(q^2) + 8.4690 \ \overline{\alpha}^3(q^2) \right)
\nonumber \\
&\times& \left(1 + 0.0208 \ \overline{\alpha}(\mu^2) + 0.0095 \ 
\overline{\alpha}^2(\mu^2)+ 0.0164 \ \overline{\alpha}^3(\mu^2) \right)
\ .
\eeq
Thus, we can combine both eqs.~(\ref{ap-solGl},\ref{ap-solFh}), as done in \eq{alphahNPt}, 
to obtain:
\beq\label{alphahNPt2}
\alpha_T(q^2) = \alpha^{\rm pert}_T(q^2)
\ \left(  \rule[0cm]{0cm}{0.8cm} 
1 \right. &+& \left. \frac{9}{q^2} 
\frac{\overline{g}^2(\mu^2) \langle A^2 \rangle_{R,\mu^2}}{4 (N_C^2-1)}
\left( \frac{\overline{\alpha}(q^2)}
{\overline{\alpha}(\mu^2)}\right)^{\frac{27}{116}} 
\right.
\nonumber \\
&\times&
\left(1 +  1.5129 \ \overline{\alpha}(q^2) + 3.8552 \ 
\overline{\alpha}^2(q^2) + 12.4952 \ \overline{\alpha}^3(q^2) \right)
\nonumber \\
&\times& \left. 
\left(1 + 0.0208 \ \overline{\alpha}(\mu^2) + 0.0095 \ 
\overline{\alpha}^2(\mu^2)+ 0.0164 \ \overline{\alpha}^3(\mu^2) \right)
\rule[0cm]{0cm}{0.8cm}
\right) \ .
\eeq
Finally, provided that the purely perturbative part of the OPE formula in 
\eq{alphahNPt2} is the coupling renormalized in the T-scheme, it seems more 
appropriate to expand in terms of $\alpha_T$ instead of $\overline{\alpha}$. 
Then, one can apply eqs.~(\ref{alpha2},\ref{cis}) for the conversion and obtain:
\beq\label{alphahNPt3}
\alpha_T(q^2) = \alpha^{\rm pert}_T(q^2)
\ \left(  \rule[0cm]{0cm}{0.8cm} 
1 \right. &+& \left. \frac{9}{q^2} 
\frac{g_T^2(\mu^2) \langle A^2 \rangle_{R,\mu^2}}{4 (N_C^2-1)}
\left( \frac{\alpha_T(q^2)}
{\alpha_T(\mu^2)}\right)^{\frac{27}{116}} 
\right.
\nonumber \\
&\times&
\left(1 +  1.2932 \ \alpha_T(q^2) + 1.9363 \ 
\alpha_T^2(q^2) + 3.8296 \ \alpha_T^3(q^2) \right)
\nonumber \\
&\times& \left. 
\left(1 - 0.7033 \ \alpha_T(\mu^2) - 0.3652 \ 
\alpha_T^2(\mu^2) + 0.0051 \ \alpha_T^3(\mu^2) \right)
\rule[0cm]{0cm}{0.8cm}
\right) \ .
\eeq
%


\end{document}